\documentclass[11pt]{article}

\usepackage[bookmarks=true,breaklinks=true]{hyperref}
\hypersetup{breaklinks=true} 

\let\fn\footnote
\renewcommand{\footnote}[1]{\linespread{1.1}\fn{#1}\linespread{1.29}}

\usepackage{fancyhdr}
\usepackage[left]{lineno}

\makeatletter\renewcommand{\section}{\@startsection
{section}{1}{\z@}{-3.5ex plus -1ex minus
    -.2ex}{2.3ex plus .2ex}{\bf }}
\makeatletter\renewcommand{\subsection}{\@startsection{subsection}{2}{\z@}{-3.25ex
plus -1ex minus
   -.2ex}{1.5ex plus .2ex}{\it }}
\makeatletter\renewcommand{\subsubsection}{\@startsection{subsubsection}{3}{-2.45ex}{-3.25ex
plus -1ex minus -.2ex}{1.5ex plus .2ex}{\it }}
\renewcommand{\thesection}{\arabic{section}.}
\renewcommand{\thesubsection}{\arabic{section}.\arabic{subsection}.}

\makeatletter\renewcommand{\subsubsection}
{\@startsection{subsubsection}{3}{\z@}{-3.25ex plus -1ex minus -.2ex}
{1.5ex plus .2ex}{\noindent\underline}}

\renewcommand{\theequation}{\thesection\arabic{equation}}
\makeatletter \@addtoreset{equation}{section}

\newcommand{\acknowledgements}{\section*{Acknowledgements}
\addcontentsline{toc}{section}{\hspace{0.6cm}{\bf Acknowledgements}}}

\newcommand{\appendices}{\section*{Appendices}\setcounter{subsection}{0}\setcounter{equation}{0}\renewcommand{\thesubsection}{\Alph{subsection}.}
\renewcommand{\theequation}{\thesubsection\arabic{equation}}
\makeatletter \@addtoreset{equation}{subsection}
\addtocontents{toc}{\vspace{0.2cm}

{\bf Appendices}}
}

\usepackage{epsfig,rotating}
\usepackage{amsmath,amssymb}
\usepackage{amsfonts}
\usepackage[matrix,arrow]{xy}

\usepackage{mathrsfs}
\usepackage{bbm}
\usepackage{bm}
\usepackage[vcentermath,enableskew]{youngtab}

\def\slasha#1{\setbox0=\hbox{$#1$}#1\hskip-\wd0\hbox to\wd0{\hss\sl/\/\hss}}

\def\periodb#1{\setbox0=\hbox{$#1$}#1\hskip-\wd0\hbox to\wd0{-}}
\newcommand{\pr}{$\,^\prime$}

\newcommand{\bfah}{\hat{\mathbf{a}}}				

\newcommand{\lsc}{\{\![}			
\newcommand{\rsc}{]\!\}}

\newcommand{\unit}{\mathbbm{1}}   			
\newcommand{\id}{\mathrm{id}}   			

\newcommand{\CA}{\mathcal{A}}    			

\newcommand{\CC}{\mathcal{C}}

\newcommand{\CD}{\mathcal{D}}
\newcommand{\CCD}{\mathscr{D}}

\newcommand{\CCH}{\mathscr{H}}

\newcommand{\CM}{\mathcal{M}}
\newcommand{\CN}{\mathcal{N}}
\newcommand{\CO}{\mathcal{O}}

\newcommand{\CV}{\mathcal{V}}

\newcommand{\frg}{\mathfrak{g}}				

\newcommand{\frh}{\mathfrak{h}}

\newcommand{\FR}{\mathbbm{R}}     			
\newcommand{\FC}{\mathbbm{C}}     			
\newcommand{\NN}{\mathbbm{N}}     			
\newcommand{\RZ}{\mathbbm{Z}}     			
\newcommand{\CPP}{{\mathbbm{C}P}}    			
\newcommand{\ah}{\hat{a}}
\newcommand{\Ah}{\hat{A}}
\newcommand{\fh}{\hat{f}}

\newcommand{\xh}{\hat{x}}
\newcommand{\dd}{\mathrm{d}}     			
\newcommand{\dpar}{\partial}     			
\newcommand{\embd}{{\hookrightarrow}}     		
\newcommand{\diag}{{\mathrm{diag}}}     		
\newcommand{\ch}{{\mathrm{ch}}}     		
\newcommand{\di}{\mathrm{i}}     			
\newcommand{\eps}{{\varepsilon}}			

\newcommand{\bz}{{\bar{z}}}

\newcommand{\bs}{{\bar{\zeta}}}

\newcommand{\eand}{{~~~\mbox{and}~~~}}     		

\newcommand{\der}[1]{\frac{\dpar}{\dpar #1}}   		
\newcommand{\derr}[2]{\frac{\dpar #1}{\dpar #2}}   	
\newcommand{\dderr}[2]{\frac{\dd #1}{\dd #2}}   	
\newcommand{\tr}{\,\mathrm{tr}\,}     			

\newcommand{\st}[2]{|#1,k\rangle\hspace{-0.18cm}\bullet\hspace{-0.18cm}\langle k,#2|}
\newcommand{\unitH}{|k\rangle\hspace{-0.18cm}\bullet\hspace{-0.18cm}\langle k|}
\newcommand{\unitHH}{|k-2\rangle\hspace{-0.18cm}\bullet\hspace{-0.18cm}\langle k-2|}

\newcommand{\aI}{\mathfrak{I}}
\newcommand{\aT}{\mathfrak{T}}

\newcommand{\au}{\mathfrak{u}}
\newcommand{\asu}{\mathfrak{su}}
\newcommand{\aso}{\mathfrak{so}}

\newcommand{\sSU}{\mathsf{SU}}
\newcommand{\sSL}{\mathsf{SL}}
\newcommand{\sGL}{\mathsf{GL}}

\newcommand{\sSO}{\mathsf{SO}}

\newcommand{\sSpin}{\mathsf{Spin}}
\newcommand{\sEnd}{\mathsf{End}\,}

\newcommand{\acton}{\vartriangleright}     			
\newcommand{\remark}[1]{}     				
\def\tyng(#1){\hbox{\tiny$\yng(#1)$}}			
\def\tyoung(#1){\hbox{\tiny$\young(#1)$}}			

\newcommand{\mbf}[1]{{\boldsymbol {#1} }}

\begin{document}

\begin{titlepage}
\begin{flushright}
 HWM--10--3 \\ EMPG--10--03
\end{flushright}
\vskip 2.0cm
\begin{center}
{\LARGE \bf Quantized Nambu-Poisson Manifolds\\[3mm] and $\boldsymbol{n}$-Lie Algebras} \vskip 1.5cm
{\Large Joshua DeBellis, Christian S\"amann, Richard J. Szabo}
 \setcounter{footnote}{0}
\renewcommand{\thefootnote}{\arabic{thefootnote}} \vskip 1cm {\em Department of Mathematics\\
Heriot-Watt University\\
Colin Maclaurin Building, Riccarton, Edinburgh EH14 4AS, U.K.\\
and Maxwell Institute for Mathematical Sciences, Edinburgh,
U.K.}\\[5mm] {E-mail: {\ttfamily jd111@hw.ac.uk , C.Saemann@hw.ac.uk , R.J.Szabo@hw.ac.uk}} \vskip 1.1cm
\end{center}
\vskip 1.0cm
\begin{center}
{\bf Abstract}
\end{center}
\begin{quote}
We investigate the geometric interpretation of quantized Nambu-Poisson
structures in terms of noncommutative geometries. We describe an
extension of the usual axioms of quantization in which classical
Nambu-Poisson structures are translated to $n$-Lie algebras at quantum
level. We demonstrate that this generalized procedure matches an
extension of Berezin-Toeplitz quantization yielding quantized spheres,
hyperboloids, and superspheres. The extended
Berezin quantization of spheres is closely related to a deformation
quantization of $n$-Lie algebras, as well as the approach based on
harmonic analysis. We find an interpretation of Nambu-Heisenberg
$n$-Lie algebras in terms of foliations of $\FR^n$ by fuzzy spheres,
fuzzy hyperboloids, and noncommutative hyperplanes. Some applications
to the quantum geometry of branes in M-theory are also briefly
discussed.
\end{quote}
\end{titlepage}


\section{Introduction and summary of results}

Modifications and extensions of classical geometry have appeared on many occasions within string theory and one of the most prominent such extensions is noncommutative geometry. A noncommutative space is defined in terms of an algebra of functions which, roughly speaking, arises by replacing the coordinate functions with noncommuting operators. This algebra corresponds in many cases of interest to the universal enveloping algebra of a Lie algebra. For example, the noncommuting coordinates $\xh^\mu$ on noncommutative euclidean space $\FR^n_\theta$ satisfy a Weyl algebra $[\xh^\mu,\xh^\nu]=\di\,\theta^{\mu\nu}$, $[\theta^{\mu\nu},\hat x^\lambda]=0$. Another well-known example is the fuzzy or Berezin-quantized sphere~\cite{Berezin:1974du}, where the operators $\hat x^\mu$ corresponding to the euclidean coordinates $x^\mu$ which satisfy $x^\mu \, x^\mu=1$, and thus describe the embedding $S^2\, \embd\, \FR^3$, form the generators of $\asu(2)$, $[\xh^\mu,\xh^\nu]=\di\,\eps^{\mu\nu\kappa}\, \xh^\kappa$.

This fuzzy sphere arises very naturally in the description of D1-branes ending on D3-branes in Type~IIB superstring theory~\cite{Myers:1999ps,Constable:1999ac} and the effective dynamics of this system is described by the Nahm equations~\cite{Diaconescu:1996rk,Tsimpis:1998zh}. Consider a static D-brane configuration consisting of a stack of D1-branes suspended between two D3-branes. Solutions to the Nahm equations factorize into a function living on the interval bounded by the D3-branes and three $\sGL(n,\FC)$-valued constants forming generators of $\asu(2)$. Geometrically, this means that the worldvolume of the D1-branes polarizes into a fibration of fuzzy spheres over the interval. 

In order to find an appropriate description of the lift of this configuration to M-theory, one can study supergravity solutions describing M2-branes ending on M5-branes. Here one realizes that the Lie algebra appearing in the original Nahm equations has to be replaced with a generalization involving ternary brackets in the lifted Nahm equations. This observation led Basu and Harvey to suggest such lifted Nahm equations based on 3-Lie algebras~\cite{Basu:2004ed}. The general concept of an $n$-Lie algebra, i.e.\ a vector space endowed with a totally antisymmetric $n$-ary bracket satisfying a generalization of the Jacobi identity known as the {\em fundamental identity}, had been introduced before by Filippov~\cite{Filippov:1985aa}. 

The desired geometric interpretation of stationary solutions to the Basu-Harvey equations is quite clear. The worldvolume of a stack of M2-branes suspended between two M5-branes should polarize into a fibration of fuzzy 3-spheres over the interval bounded by the two M5-branes. The identification of these fuzzy 3-spheres with the ones suggested by Guralnik and Ramgoolam~\cite{Guralnik:2000pb}, however, does not quite match, see e.g.~\cite{Nastase:2009ny}. More importantly, there is no known generalization of any of the frameworks of noncommutative geometry which yields operator algebras built on $n$-Lie algebras.

In this paper we study this problem in detail. In particular, we will construct and identify many noncommutative spaces whose algebra of functions can be endowed with an $n$-Lie algebra bracket. We start from the canonical axioms of quantization and attempt to find suitable generalizations for each one of them. Here the problematic axiom is the correspondence principle, which states that the commutator of two quantum operators is proportional to the quantization of the Poisson bracket of the corresponding classical observables. One is immediately led to replacing the commutator with the $n$-bracket of an $n$-Lie algebra. Nambu brackets~\cite{Nambu:1973qe,Takhtajan:1993vr} provide a natural extension of Poisson brackets to an $n$-ary bracket. They endow the vector space of smooth functions on a manifold with an $n$-Lie algebra structure. In addition to the fundamental identity of $n$-Lie algebras, they satisfy a generalized Leibniz rule as well. 

Nambu's original aim was to define an extended hamiltonian mechanics built on these brackets. For consistency of the dynamics, both the fundamental identity and the generalized Leibniz rule are required. This makes the quantization of Nambu mechanics notoriously difficult, as the only known example\footnote{There is, however, the very formal exception of Zariski quantization, which yields a consistent quantization of Nambu mechanics~\cite{Dito:1996xr,Dito:1996hn}.} of a vector space endowed with an $n$-ary bracket with $n>2$ meeting the consistency conditions is the classical Nambu-Poisson algebra. In particular, it is not sufficient to have an operator algebra forming an $n$-Lie algebra at quantum level. 

Recall, however, that it is not our aim to solve the problem of quantizing Nambu mechanics but merely to find geometric interpretations of operator algebras in terms of quantized algebras of functions which are endowed with an $n$-Lie bracket. That is, we just solve the kinematical problem of quantizing Nambu mechanics, which consists of providing a quantization prescription mapping classical observables to quantum operators. We do not solve the dynamical problem of deriving quantum dynamics from the classical Nambu mechanics. For this reason, the correspondence principle, which guarantees a solution to the dynamical problem of quantization in geometric quantization, will here merely serve as a guiding principle and it plays a much less fundamental role than the binary operation on the operator algebra.

The quantization prescriptions used in this paper are extensions of Berezin-Toeplitz quantization, a mixture of geometric quantization and deformation quantization. While it makes use of the Hilbert space of geometric quantization, it requires the correspondence principle only to be satisfied to first order in a discrete deformation parameter $\hbar$. One thus starts from a suitable complex line bundle $L$ over a K{\"a}hler manifold $\CM$ and identifies the quantum Hilbert space $\CCH_L$ with the vector space of global holomorphic sections of $L$. Classical observables, i.e.\ functions on $\CM$, are mapped to endomorphisms of $\CCH_L$. 

We extend this quantization procedure to various Nambu-Poisson manifolds, in particular to spheres, and give the explicit quantization map. Our approach is based on an embedding of general hyperboloids into complex projective space, which is naturally provided by generators of certain Clifford algebras. Although this construction is not intrinsic, recall that the line bundle used in Berezin-Toeplitz quantization is very ample, implying the existence of an embedding of $\CM$ into complex projective space by the Kodaira embedding theorem\footnote{In fact, any choice of basis for the quantum Hilbert space $\CCH_L$ here provides such an embedding.}. From this point of view, the approach via an embedding is very natural.

To endow the operator algebra with an $n$-Lie bracket, we use a recently proposed truncation of the classical Nambu-Poisson bracket~\cite{Ho:2008bn,Chu:2008qv}, which yields an $n$-Lie algebra structure on the set of polynomials truncated at a certain maximal degree. Using the quantization map, we can lift this $n$-Lie algebra structure to the operator algebra and the generalized correspondence principle is satisfied, essentially by definition. In some cases of interest, e.g.\ for spheres $S^d$ with $d\leq 4$, this bracket turns out to be equivalent to the $n$-Lie bracket given by the totally antisymmetric operator product at linear level. It is also only at linear level that this $n$-Lie algebra structure reduces to the ordinary Dirac quantization prescription for $n=2$. We interpret the latter as one of many hints that a quantization involving Nambu-Poisson and $n$-Lie algebra structures should be performed in a different way. Nevertheless, the operator algebras we find by our method should suffice for the physical applications to branes in M-theory that sparked our investigation.

One such alternative approach would be to look for a geometric
interpretation of the multisymplectic $d$-form $\varpi$ corresponding
to the Nambu bracket on $S^d$. In the same way that the symplectic
2-form on $S^2$ encodes a line bundle, the multisymplectic
3-form on $S^3$ corresponds to a gerbe. Roughly speaking, one
would then arrive at a quantization of the infinite-dimensional
K{\"a}hler manifold corresponding to the loop space of $S^3$. This was
also suggested in~\cite{Takhtajan:1993vr}, where an action principle
based on 2-chains for Nambu mechanics was proposed, and it naturally
emerges in certain limits of M-theory in constant $C$-field
backgrounds from the quantization of open membranes ending on
M5-branes~\cite{Bergshoeff:2000jn,Kawamoto:2000zt}. Since it seems
that many of the necessary technical details still remain to be worked
out\footnote{Even basic aspects, e.g. the fact that a 3-form is not enough to define a Nambu bracket, remain to be put into context.} and since we would like to end up with finite-dimensional Hilbert spaces for physical reasons, we choose not to pursue this approach here.

We also introduce the notion of a universal enveloping algebra of an $n$-Lie algebra and find that the usual quantization prescription using universal enveloping algebras agrees with our generalized Berezin-Toeplitz procedure. Moreover, there is a close relationship between our quantization prescription applied to spheres and the noncommutative spheres which have appeared in the literature so far, see~\cite{Grosse:1996mz,Guralnik:2000pb,Ramgoolam:2001zx} as well as~\cite{Medina:2002pc,Dolan:2003kq,Abe:2004sa}. As we allow for radial fuzziness of our quantized spaces, which was eliminated in~\cite{Guralnik:2000pb,Ramgoolam:2001zx}, our quantum 3-sphere does form suitable solutions to the Basu-Harvey equations. We also derive explicit formulas for the Laplace operators as well as integrals over the quantum spheres. 

We can easily adjust the Clifford algebra appearing in our quantization prescription of spheres to Clifford algebras arising from pseudo-riemannian metrics. This procedure directly yields quantum hyperboloids, and in particular fuzzy de~Sitter and anti-de~Sitter spaces are readily constructed. For this, one has to sacrifice the quantization axiom that observables are mapped to hermitian operators, but as long as we are only concerned with the kinematical problem of quantization, this is not an obstacle. Similarly, we extend our construction to superspheres.

We will also describe the geometric interpretation of Nambu-Heisenberg $n$-Lie algebras, which are generalizations of the Heisenberg algebra. We find that to pin down the actual geometry, one should introduce further Nambu-Poisson structures with brackets of lower degrees. One then obtains an interpretation of these multi-Nambu-Poisson structures as 
a quantum euclidean space $\FR^n$ which is foliated by quantum hyperplanes or quantum spheres. The special case $\FR^3_\lambda$, which is a discrete foliation of $\FR^3$ by fuzzy spheres, was previously studied e.g.\ in~\cite{Hammou:2001cc,Batista:2002rq}. In this manner, we interpret solutions to an equation recently found by Chu and Smith~\cite{Chu:2009iv} in the description of M2-branes ending on M5-branes in a background $C$-field. The polarized worldvolume of the M2-branes corresponds to the noncommutative space $\FR^{1,2}_{\lambda'}\times \FR^3_\lambda$.

There is a vast amount of literature concerned with problems related to those discussed in this paper. For further reading, see e.g.~\cite{Hoppe:1996xp,Awata:1999dz,Minic:2002pd,Curtright:2002sr,Kawamura:2002yz,Curtright:2002fd,Axenides:2008rn,Curtright:2009qf} and references therein.

This paper is structured as follows. We review Nambu-Poisson structures and $n$-Lie algebras as well as the appearance of the latter in recent developments in M-theory in Section~2. The generalization of the canonical axioms of quantization to the case of Nambu-Poisson structures is developed in Section~3, where we also discuss the quantization approach via universal enveloping algebras. Section~4 gives a brief summary of Berezin-Toeplitz quantization, which is extended in Section~5 to the case of spheres. The further extensions to hyperboloids and superspheres are given in Sections~6 and~7. In Section~8, we discuss the analogous quantization of $\FR^n$ which yields foliations in terms of noncommutative spheres and hyperplanes. Five appendices at the end of the paper contain some of the more technical details and results of our calculations.

\section{Extensions of Poisson and Lie brackets}

The Nambu $n$-bracket is an extension of the Poisson bracket to a
bracket acting on $n$ functions, satisfying both a generalized Leibniz
rule and a generalized Jacobi identity. Similarly, an $n$-Lie algebra
is an extension of Lie algebras built on a bracket with $n$ slots
which only satisfies the generalized Jacobi identity. These structures play a prominent role in recent proposals for descriptions of M-brane configurations. We briefly review these $n$-ary brackets in this section.

\subsection{Nambu brackets}

A {\em Nambu-Poisson structure}~\cite{Nambu:1973qe,Takhtajan:1993vr}
on a smooth manifold $\CM$ is an $n$-ary, totally antisymmetric linear map $\{-,\dots ,-\}:\CC^\infty(\CM)^{\wedge n}\rightarrow\CC^\infty(\CM)$, which satisfies the {\em generalized Leibniz rule}
\begin{equation}
 \{f_1 \,f_2,f_3,\dots ,f_{n+1}\}=f_1\,\{f_2,\dots
 ,f_{n+1}\}+\{f_1,\dots ,f_{n+1}\} \,f_2
\end{equation}
as well as the {\em fundamental identity}
\begin{equation}
 \{f_1,\dots ,f_{n-1},\{g_1,\dots ,g_n\}\}=\{\{f_1,\dots ,f_{n-1},g_1\},\dots ,g_n\}+\dots +\{g_1,\dots ,\{f_1,\dots ,f_{n-1},g_n\}\}
\end{equation}
for $f_i,g_i\in\CC^\infty(\CM)$. 
The map $\{-,\dots ,-\}$ is called a {\em Nambu $n$-bracket}, the
manifold $\CM$ is called a {\em Nambu-Poisson manifold}, and we call the algebra of smooth functions $\CC^\infty(\CM)$ endowed with the Nambu $n$-bracket a {\em Nambu-Poisson algebra}.
The Leibniz rule and the fundamental identity imply that the manifold $\CM$
admits an $n$-vector field $\varpi\in (T\CM)^{\wedge n}$ called a {\em Nambu-Poisson tensor}, such that
\begin{equation}
\{f_1,\dots ,f_n\}=\varpi (\dd f_1\wedge \dots \wedge \dd f_n)
\end{equation}
for all $f_i \in \CC^\infty(M)$.

In this paper we will be predominantly interested in the case where
$\CM$ is a sphere. Recall that the canonical symplectic structure on the sphere $S^2$ reads as
\begin{equation}
 \omega=\left(\begin{array}{cc} 0 & {\rm vol}_\theta \\ -{\rm vol}_\theta &0\end{array}
\right)
\end{equation}
in the basis given by the usual angular coordinates
$\varphi=(\varphi^1,\varphi^2):=(\theta,\phi)$, where
$\theta\in[0,\pi]$ and $\phi\in[0,2\pi]$. Here ${\rm vol}_\theta=\sin
\theta$ is the volume element on $S^2$. The $2$-vector field $\varpi$
defining the Poisson or Nambu $2$-bracket is obtained by inverting the
matrix $\omega$, and we have\footnote{Throughout this paper, we will
  always implicitly sum over repeated indices irrespective of their positions.}
\begin{equation}
 \{f_1,f_2\}:=\varpi(\dd f_1\wedge \dd f_2)=\frac{\eps^{ij}}{{\rm
     vol}_\theta} \,\derr{f_1}{\varphi^i} \, \derr{f_2}{\varphi^j}~.
\end{equation}

Analogously, we define the $d$-vector field $\varpi$ yielding the Nambu
$d$-bracket on $S^d$ parameterized by the usual angular coordinates
$\varphi^i$ by
\begin{equation}
 \{f_1,\ldots,f_d\}:=\varpi (\dd f_1\wedge \dots \wedge \dd
 f_d):=\frac{\eps^{i_1\ldots i_d}}{{\rm vol}_{\varphi}} \, \derr{f_1}{\varphi^{i_1}}\ldots\derr{f_d}{\varphi^{i_d}}~.
\end{equation}
Consider now the standard embedding of the sphere $S^d$ of radius $R$ into $\FR^{d+1}$, where the cartesian coordinates $x^\mu$, $\mu=1,\dots ,d+1$ are given by
\begin{equation}\label{sphere.coordinates}
 x^1=R\, \cos(\varphi^1)~,~~~x^2=R\, \sin(\varphi^1)\,\cos(\varphi^2)~,~~~x^3=R\, \sin(\varphi^1) \, \sin(\varphi^2) \,\cos(\varphi^3)~, ~~~\ldots~.
\end{equation}
This embedding induces the volume element on $S^d$ given in spherical
coordinates by
\begin{equation}\label{sphere.vol}
{\rm vol}_\varphi:=R^d\, \sin^{d-1}(\varphi^1) \, \sin^{d-2}(\varphi^2)\cdots
\sin(\varphi^{d-1})~.
\end{equation}
We will not use ${\rm vol}_\varphi$ directly in the definition, but rescale it by a factor of $R^{1-2d}$. The Nambu $d$-bracket of the embedding coordinate functions $x^\mu(\varphi^i)$ is then readily calculated to be
\begin{equation}\label{NPbracketPolynomial}
 \big\{x^{\mu_1}(\varphi^i),\ldots,x^{\mu_{d}}(\varphi^i)
 \big\}=R^{d-1}\, \eps^{\mu_1\ldots\mu_d\mu_{d+1}}\,x^{\mu_{d+1}}(\varphi^i)~.
\end{equation}
One can extend this bracket to polynomials in $x^\mu$ by using the generalized Leibniz rule as shown in Appendix \ref{ExtNPalgebra} These polynomials in turn span the space of hyperspherical
harmonics, as we discuss later on. The bracket \eqref{NPbracketPolynomial} is naturally invariant
under the isometry group $\sSO(d+1)$ of $S^d$.

\subsection{$n$-Lie algebras}

An {\em $n$-Lie algebra}~\cite{Filippov:1985aa} is a vector space $\CA$ equipped with a totally antisymmetric, multilinear bracket $[-,\dots ,-]: \CA^{\wedge n}\rightarrow \CA$, which satisfies the {\em fundamental identity}
\begin{equation}
\big[x^1,x^2,\dots ,x^{n-1},[y^1,y^2,\dots ,y^n]\big]=\sum^{n}_{i=1}
\, \big[y^1,\dots ,[x^1,\dots , x^{n-1},y^i],\dots ,y^n\big]
\label{fundid}\end{equation}
for all $x^i$, $y^i\in \CA$. In particular, the algebra of smooth functions $\CC^\infty(\CM)$ of a Nambu-Poisson manifold $\CM$ forms an infinite-dimensional $n$-Lie algebra with the $n$-Lie bracket given by the Nambu $n$-bracket.

The fundamental identity is a generalization of the
Jacobi identity. While the adjoint action of a Lie algebra on itself
generates its inner derivations, the space of inner derivations of an
$n$-Lie algebra $\CA$ is spanned by operators $D(x^1 \wedge\dots
\wedge x^{n-1})\in\mathfrak{gl}(\CA)$, $x^i\in\CA$, defined by
\begin{align}
D(x^1 \wedge\dots \wedge x^{n-1})\cdot y:=[x^1,\dots ,x^{n-1},y]
\end{align}
for $y\in\CA$. The inner derivations form a Lie algebra
\begin{equation}
 \big[D(x),D(y)\big]\cdot z:=D(x)\cdot\big(D(y)\cdot
 z\big)-D(y)\cdot\big(D(x)\cdot z \big)~,~~~x,y\in \CA^{\wedge (n-1)}~,~z\in\CA~,
\end{equation}
where closure of the Lie bracket is guaranteed by the fundamental identity. We call the Lie algebra of inner derivations of an $n$-Lie algebra $\CA$ its {\em associated Lie algebra}~$\frg_\CA$.

One can reduce an $n$-Lie algebra $\CA$ to an $n-1$-Lie
algebra $\CA'$, cf.~\cite{Filippov:1985aa}. One chooses an element
$x_0\in\CA$ and identifies the vector space of $\CA'$ with $\CA$. The $n-1$-Lie
bracket on $\CA'$ is defined as $[x^1,\dots
,x^{n-1}]_{\CA'}=[x^1,\dots ,x^{n-1},x_0]$,
$x^i\in\CA$. By placing an inner product on the vector space $\CA'$, we can moreover restrict $\CA'$ to the orthogonal complement of $x_0$ in $\CA'$. Applying this procedure $n-2$ times, we arrive at a
second Lie algebra $\frh_\CA$ starting from $\CA$, which generally differs from $\frg_\CA$.

Let us give a class of explicit examples which are abstractions of the
Nambu-Poisson algebras for spheres. The $d$-Lie algebra $A_{d+1}$~\cite{Filippov:1985aa} is given by a $d+1$-dimensional $\FC$-vector space with basis $(x^1,\dots ,x^{d+1})$ endowed with the $d$-bracket
\begin{equation}
{}[x^{i_1},\dots,x^{i_d}]=\eps^{i_1\dots i_di_{d+1}} \,x^{i_{d+1}} ~.
\end{equation}
Its associated Lie algebra $\frg_{A_{d+1}}$ is $\aso(d+1)$, and the
$2$-Lie algebra $\frh_{A_{d+1}}$ obtained by reducing with respect to $d-2$ arbitrary
elements is $\aso(3)\cong\asu(2)$. This is the unique simple $d$-Lie
algebra over $\FC$, cf.~e.g.~\cite{Figueroa-O'Farrill:arXiv0805.4760} and references therein. The case $d=3$, for which
$\frg_{A_4}=\aso(4)\cong\asu(2)\oplus \asu(2)$, is the most prominent
3-Lie algebra in the context of the recently conjectured multiple
M2-brane gauge theories~\cite{Bagger:2006sk,Gustavsson:2007vu}.

\subsection{Truncation of Nambu-Poisson brackets}\label{sec.truncNambu}

Let us assume that the components of the Poisson tensor $\varpi$ on a smooth manifold $\CM$ are given by homogeneous polynomials of degree $d(\varpi)\geq 1$ in some coordinates $(x^\mu)$. If the polynomial ring $\FC[x^\mu]$ is furthermore a subset of $\CC^\infty(\CM)$, then there is a truncation of the Nambu-Poisson algebra $\CC^\infty(\CM)$ to an $n$-Lie algebra structure on $\FC[x^\mu]$ \cite{Ho:2008bn,Chu:2008qv} as reviewed below.

We define for every $K\in \NN$ a totally antisymmetric, linear $n$-bracket on $\FC[x^\mu]$ according to
\begin{equation}
 \{f_1,\ldots,f_n\}_K:=\left\{\begin{array}{ll}
\{f_1,\ldots,f_n\}&\mbox{if}~~d(f_1)+\ldots+d(f_n)+d(\varpi)-n\leq K\\
0&\mbox{else}
\end{array}
\right.~,
\end{equation}
where $f_i\in\FC[x^\mu]$ and $d(f_i)$ denotes the degree of the polynomial $f_i$. It is immediately clear that the Leibniz rule cannot survive the truncation. The fundamental identity, however, does, as we show in the following, cf.\ \cite{Ho:2008bn,Chu:2008qv}. Let $f_i,g_i\in \FC[x^\mu]$. We then have
\begin{equation}\label{checkFI}
 \{f_1,\ldots,f_{n-1},\{g_1,\ldots,g_n\}_K\}_K=\sum_{i=1}^n \{g_1,\ldots,\{f_1,\ldots,f_{n-1},g_i\}_K,\ldots,g_n\}_K~.
\end{equation}
The cases $d(f_i)=0$ or $d(g_i)=0$ for some $i$ are trivial, let us therefore assume that $d(f_i)>0$ and $d(g_i)>0$. Equation \eqref{checkFI} is nontrivial if and only if the outer brackets on either side are non-vanishing, which amounts to
\begin{equation}
 d(f_1)+\ldots+d(f_{n-1})+d(g_1)+\ldots+d(g_n)+2d(\varpi)-2n\leq K~.
\end{equation}
Because of $d(\varpi)\geq 1$, it is easy to see that this condition also implies that none of the inner brackets of \eqref{checkFI} vanish. Thus, whenever \eqref{checkFI} is nontrivial, the brackets are given by the ordinary Nambu-Poisson brackets and thus satisfy the fundamental identity.

\subsection{Generalized Nahm equations}

In Type~IIB string theory, magnetic monopoles of charge $N$ can be regarded as a stack of $N$ D1-branes ending on a D3-brane~\cite{Callan:1997kz}. From the perspective of the D1-brane string theory, the effective dynamics is described by the Nahm equations
\begin{equation}
\dderr{T^i}{s}+\eps^{ijk}\,[T^j,T^k]=0 \ ,
\end{equation}
where $T^i$ describe fluctuations of the D1-branes parallel to the worldvolume of the D3-brane. These equations have a solution $T^i(s)=f(s)\,\tau^i$, where $f(s)=\frac1s$ and $\tau^i=\eps^{ijk}\,[\tau^j,\tau^k]$, which describes the transverse scalar fields by a fuzzy 2-sphere~\cite{Myers:1999ps,Constable:1999ac}. The two extra fuzzy dimensions are required to reconstruct the D3-brane from the D1-branes.

The Basu-Harvey equations~\cite{Basu:2004ed} conjecturally describe stacks of M2-branes ending on an M5-brane in M-theory, analogously to the Nahm equations describing stacks of D1-branes ending on a D3-brane. Suitably reformulated, they read
\begin{equation}
 \dderr{T^i}{s}+\eps^{ijkl}\, [T^j,T^k,T^l]=0~.
\end{equation}
It should allow for a solution via factorization $T^i(s)=f(s)\,\tau^i$, where $f(s)=\frac{1}{\sqrt{2s}}$ and $\tau^i=\eps^{ijkl}\,[\tau^j,\tau^k,\tau^l]$. Thus the transverse scalar fields $T^i$ could live in the 3-Lie algebra $A_4$, which describes the intersecting configuration in terms of multiple M2-branes again as a fuzzy funnel, this time with the extra three worldvolume dimensions of the M5-brane arising as a fuzzy 3-sphere. The associated Lie algebra is $\aso(4)$, which corresponds to the correct group of isometries of $S^3$. Fixing one slot reduces $A_4$ to $\asu(2)$, and describes the isometries in the reduction of the M-brane system to the D-brane system above.

In~\cite{Chu:2009iv} it was demonstrated how the Nahm equations can be understood as a boundary condition for open strings. This point of view becomes particularly fruitful when examining how the worldvolume geometry of the D3-brane is deformed by a constant $B$-field applied in the transverse directions to the D1-branes. This induces a constant shift in the Nahm equations which can be accounted for by a noncommutative geometry on the D3-brane, described by the Heisenberg commutation relations
\begin{equation}
[T^i,T^j]=\di\,\theta^{ij} \ ,
\label{D3NCG}\end{equation}
where $\theta^{ij}$ is a constant antisymmetric matrix whose components are related to the components of the $B$-field.

Analogously, the Basu-Harvey equations can be derived as a boundary condition of open membranes. By including a constant $C$-field on the M5-brane, one can reproduce the M2-brane funnel from the M5-brane point of view if the Basu-Harvey equations are suitably modified~\cite{Chu:2009iv}. This modification identifies the open membrane boundary conditions in the presence of a $C$-field, which describes the M5-brane worldvolume by a quantum geometry of the form
\begin{equation}
[T^i,T^j,T^k]=\di\,\Theta^{ijk} \ ,
\label{M5NCG}\end{equation}
where $\Theta^{ijk}$ is a totally antisymmetric constant tensor whose components are related to the components of the constant $C$-field.

Both the Nahm equations and the Basu-Harvey equations are special cases of generalized Nahm equations built on $n$-Lie algebras. These equations in turn are the homotopy Maurer-Cartan equations of special $L_\infty$ or strong homotopy Lie algebras, see~\cite{Lazaroiu:2009wz} for details. Just like the commutator (\ref{D3NCG}) arises by quantizing a Poisson bracket on $\FR^2$, it is suggested in~\cite{Chu:2009iv} that the correct form of the 3-Lie algebra (\ref{M5NCG}) is given by a quantization of the Nambu 3-bracket on $\FR^3$, see e.g.~\cite{Ho:2009zt} and references therein. The purpose of this paper is to describe various aspects of the quantization of the sorts of $n$-Lie algebras encountered above. We will find in fact that all the structures described above are naturally interlaced with each other in the quantization of certain manifolds.

\section{Quantization of Nambu-Poisson structures\label{NPquant}}

In this section we describe the general quantization procedure which
we shall employ in this paper, as a generalization of the canonical
one. We will describe this approach from both a geometric and an algebraic perspective. For the quantum geometries which we will consider, these prescriptions will be equivalent.

\subsection{Conventional quantization}

The problem of quantization splits into two parts. The first is to establish the kinematical relationship between classical and quantum observables. The second is to deduce the dynamical laws of a quantum system from their classical counterparts.

Classically, the state space of a dynamical system is a Poisson manifold $\CM$ and the observables are the smooth functions on $\CM$. One often demands that the Poisson structure is non-degenerate, which requires that $\CM$ has even dimension and turns the Poisson structure into a symplectic structure on $\CM$. At the quantum level, the states of a physical system are given by rays in a complex Hilbert space $\CCH$ and observables are linear operators acting on $\CCH$. A quantization is a map $\widehat{-}:\CC^\infty(\CM)\rightarrow \sEnd(\CCH)$, which assigns to each smooth function $f$ on $\CM$ an operator $\fh$ acting on $\CCH$.

The problem of finding a quantization for a given Poisson manifold is highly nontrivial and not understood in full generality. As an example, let us consider the special case $\CM=T^*\FR^n$ with euclidean coordinates $(x^\mu,p_\mu)$, $\mu=1,\ldots,n$, and the non-vanishing Poisson bracket $\{x^\mu,p_\nu\}=\delta^\mu{}_\nu$. One is naturally led to imposing the following axioms, which yield a {\em full quantization} (cf.\ e.g.~\cite{0821844385}):
\begin{itemize}
 \item[Q1.] The map $f\mapsto \fh$ is linear over $\FC$ and maps smooth real functions on $\CM$ to hermitian linear operators on $\CCH$.
 \item[Q2.] If $f$ is a constant function, then $\fh$ is scalar multiplication by the corresponding constant.
 \item[Q3.] The {\em correspondence principle}: If $\{f_1,f_2\}=g$ then $[\fh_1,\fh_2]=-\di\,\hbar \,\hat{g}$.
 \item[Q4.] The operators $\xh^\mu$ and $\hat{p}_\mu$ act irreducibly on $\CCH$.
\end{itemize}
Here $f,f_i,g\in \CC^\infty(\CM)$ and $\{-,-\}$ and $[-,-]$ denote the Poisson bracket on $\CM$ and the commutator of elements of $\sEnd(\CCH)$, respectively. But the Gr\"onewold-van~Howe theorem now states that there is no such quantization, see~\cite{0821844385} or~\cite{Gotay:9605001} for details. One can prove an analogous theorem for $\CM=S^2$. A full quantization of the torus $\CM=T^2$ does however exist.

There are three common loopholes to this obstruction. First, one can drop irreducibility and ignore axiom Q4. Second, one can quantize only a subclass of functions in $\CC^\infty(\CM)$. Third, one can generalize the correspondence principle such that it only holds up to first order in $\hbar$. The first two approaches lead to prequantization and further to the formalism of {\em geometric quantization}~\cite{Woodhouse:1992de}, while the third approach leads to approximate operator representations and eventually to the machinery of {\em deformation quantization}~\cite{Bayen:1977ha,Kontsevich:1997vb}. Recall also that the canonical quantization prescription of Weyl, von Neumann and Dirac is not Q3, but just the corresponding condition on the coordinates of phase space, which further supports the third approach.

All our constructions are based on Berezin\footnote{By Berezin quantization, we mean the standard constructions of fuzzy geometry. The algebra of functions is reduced to the algebra of lower Berezin symbols of $\sEnd(\CCH)$, where the product is given by the corresponding operator product.} and Toeplitz quantization, which are hybrids of geometric and deformation quantization. They both rely on the Hilbert space constructed in geometric quantization but satisfy the correspondence principle only to first order in $\hbar$. Moreover, one restricts oneself to quantizing only a subset of functions in Berezin quantization. For further details on the relations between the various approaches, see e.g.~\cite{Ali:2004ft}.

We will thus impose axioms Q1 and Q2, and axiom Q3 only to linear order in $\hbar$. If $\CM$ is a homogeneous space, we replace Q4 by
\begin{itemize}
 \item[Q4\pr.] The Hilbert space $\CCH$ carries a representation of the isometry group of $\CM$.
\end{itemize}
In Berezin-Toeplitz quantization, these representations are usually irreducible. In our extension of this construction we will, however, have to allow for reducible representations as well.

We will not require that quantizing a complete set of classical observables yields a complete\footnote{Completeness here means Schur's lemma: if an operator commutes with each element, it is proportional to the identity. Completeness in the classical case is the analogous statement involving the Poisson bracket and the constant function.} set of quantum observables, which would establish a one-to-one correspondence between $\sEnd(\CCH)$ and $\CC^\infty(\CM)$. It is well-known in geometric quantization that this would yield a Hilbert space which is too large~\cite{Woodhouse:1992de}. It is therefore necessary to endow $\CM$ with some additional structure which restricts the Hilbert space. 

In geometric quantization, a quantization map satisfying the axioms Q1--Q3 is called a {\em prequantization}. The additional condition restricting the Hilbert space is here provided by a {\em polarization} and turns a prequantization into a {\em quantization}. Recall that a polarization of a symplectic manifold $(\CM,\omega)$ is a foliation by lagrangian submanifolds, i.e.\ a smooth distribution $\CCD$ which is integrable and lagrangian in the sense that at each $p\in \CM$, $\CCD_p$ is a lagrangian subspace of $T_p\CM$ with respect to the symplectic structure $\omega_p$. A {\em complex polarization} is defined in an analogous way with respect to a complex structure on $\CM$.

\subsection{Generalized quantization axioms for Nambu brackets}

The problem of quantizing Nambu-Poisson manifolds via geometric quantization is notoriously difficult. For a discussion of these difficulties, see e.g.~\cite{Nambu:1973qe,Takhtajan:1993vr} as well as~\cite{Dito:1996xr}. Here we are merely interested in finding a quantization analogous to the Berezin-Toeplitz quantization of K{\"a}hler manifolds. In other words, we are exclusively interested in the kinematical problem of quantization. 

We start by demanding that a quantization associates to a Nambu-Poisson manifold $\CM$ a Hilbert space $\CCH$ and maps a set of {\em quantizable functions} $\Sigma\subset \CC^\infty(\CM)$ on $\CM$ to endomorphisms on $\CCH$. We impose the quantization conditions Q1, Q2, and Q4\pr, but relax Q3 in the spirit of Berezin-Toeplitz quantization. The quantization map will always be injective, and on its image $\widehat{\Sigma}\subset \sEnd(\CCH)$ we introduce its inverse $\sigma$. (In Berezin-Toeplitz quantization, $\sigma$ is the lower Berezin symbol.) The axiom Q3 is then modified to 
\begin{itemize}
 \item[Q3\pr.] The quantization maps a subalgebra of the Nambu-Poisson algebra on $\CM$ to an $n$-Lie algebra structure on a subspace of $\sEnd(\CCH)$, which satisfies the constraint\footnote{We assume the existence of a measure $\dd \mu$ on $\CM$. As we quantize K{\"a}hler manifolds exclusively, we can use the Liouville volume form $\dd \mu=\frac{\omega^n}{n!}$, where $\omega$ is the K\"ahler 2-form and $\dim_\FC \CM=n$.}
\begin{equation}
 \lim_{\hbar\rightarrow 0}\,\Big\| \frac{\di}{\hbar}\,\sigma\big([\fh_1,\ldots,\fh_n]\big)-\{f_1,\ldots,f_n\}\Big\|_{L^2}=0
\end{equation}
for all quantizable functions $f_i\in\Sigma$.
\end{itemize}
In conventional quantization, $\sigma$ is bijective and therefore the correspondence principle as stated here is equivalent to the usual one formulated in terms of operators.

The canonical choice for an $n$-ary linear and totally antisymmetric bracket on $\sEnd(\CCH)$ in the literature (cf.\ e.g.~\cite{Nambu:1973qe,Takhtajan:1993vr,Curtright:2002sr}) is the totally antisymmetric operator product
\begin{equation}
 [\fh_1,\ldots,\fh_n]:=\eps^{i_1\ldots i_n}\, \fh_{i_1}\ldots\fh_{i_n}~.
\label{antisymop}\end{equation}
This bracket neither satisfies the fundamental identity nor the Leibniz rule, in general. 

A different bracket can be defined on Nambu-Poisson manifolds, on which we can truncate the Nambu-Poisson structure as discussed in Section \ref{sec.truncNambu}: In the cases we are interested in, the set of quantizable functions $\Sigma$ is a set of polynomials of a certain maximal degree $K$. On this set, an $n$-Lie algebra structure is given by the truncated Nambu-Poisson bracket $\{-,\ldots,-\}_K$. This $n$-Lie algebra structure can be lifted from $\Sigma$ to an $n$-Lie algebra structure on $\sEnd(\CCH)$: The bracket
\begin{equation}
 {}[\Ah_1,\ldots,\Ah_n]:=\sigma^{-1}(-\di\hbar\{\sigma(\Ah_1),\ldots,\sigma(\Ah_n)\}_K)
\end{equation}
is linear, antisymmetric and satisfies the fundamental identity for arbitrary operators $\Ah_i\in\sEnd(\CCH)$, as $\sigma\circ\sigma^{-1}=\id$. 

A few remarks on this bracket are in order. First, note that for $\hbar\rightarrow 0$, we have $K\rightarrow \infty$, and the truncated $n$-Lie algebra approaches the Nambu-Poisson algebra on $\CM$. For this reason, the correspondence principle Q3\pr\ is satisfied by definition. Second, some cases this bracket will turn out to be equal to the totally antisymmetric operator product if all the arguments are linear polynomials. Third, for $n=2$, this bracket does not reproduce the commutator, but a deformation thereof.

\subsection{Deformation quantization of $n$-Lie algebras}

Noncommutative spaces whose coordinate algebras are given by the
enveloping algebras of finite-dimensional Lie algebras $\frg$ are
deformation quantizations of a linear Poisson structure, with the
universal enveloping algebra $U(\frg)$ regarded as a deformation of
the commutative (symmetric) polynomial algebra $S(\frg)$ via the
Poincar\'e-Birkhoff-Witt (PBW) theorem. We will now describe how this
generalizes to quantum geometries which are encoded in a
$d$-dimensional $n$-Lie algebra $\CA$ over $\FC$, with $n$-Lie
bracket $[-,\dots,-]:\CA^{\wedge n}\to \CA$ satisfying the fundamental
identity (\ref{fundid}) and with given basis ($x^1,\dots,x^d)$ obeying
\begin{equation}
[x^{i_1},\dots,x^{i_n}]=f^{i_1\dots i_ni_{n+1}}\,x^{i_{n+1}} \ .
\end{equation}
The totally antisymmetric structure constants $f^{i_1\dots i_{n+1}}$ are linear in
the deformation parameter $\hbar$. For the examples that we consider, this quantization will turn out to be equivalent to that provided by our generalized quantization axioms. 

Let $T(\CA)=\bigoplus_{k\geq0}\,\CA^{\otimes k}$, $\CA^0:=\FC$, be the free tensor algebra of the vector space $\CA$. Let $I_n(\CA)$ be the two-sided ideal of $T(\CA)$ generated by 
\begin{equation}
[x^1,\dots,x^n]-\varepsilon^{i_1\dots i_n}\,x^{i_1}\otimes\cdots\otimes x^{i_n}
\end{equation}
for $x^i\in\CA$. The universal enveloping algebra of $\CA$ is defined\footnote{Enveloping algebras for other sorts of ternary extensions of Lie algebras have been considered in~\cite{Goze:2008ku}.} to be the quotient $U_n(\CA):=T(\CA)/I_n(\CA)$. The fundamental identity (\ref{fundid}) in $U_n(\CA)$ is a direct consequence of associativity of the tensor product in $T(\CA)$.

The universal property of $U_n(\CA)$ may be stated as follows. Let $A$ be an associative algebra over $\FC$, and let $L_n(A)$ be the $n$-Lie algebra defined from $A$ by the $n$-Lie bracket
\begin{equation}
[a_1,\dots,a_n]_A:=\eps^{i_1\dots i_n}\,a_{i_1}\cdots a_{i_n}
\end{equation}
for $a_i\in A$. Let $f:\CA\to L_n(A)$ be a morphism of $n$-Lie algebras, i.e.\ $f:\CA\to A$ is a linear map of underlying vector spaces such that $f[x^1,\dots,x^n]=[f(x^1),\dots,f(x^n)]_A$. Then there exists a unique morphism of associative algebras $\varphi:U_n(\CA)\to A$ such that $\varphi\circ \iota_\CA=f$, where $\iota_\CA$ is the composition of the canonical injection of $\CA$ into $T(\CA)$ with the canonical surjection of $T(\CA)$ onto $U_n(\CA)$. The proof of this fact is elementary. By definition of the tensor algebra, the map $f:\CA\to L_n(A)$ extends to a morphism of associative algebras $\tilde f:T(\CA)\to A$ given by $\tilde f(x\otimes y)=f(x)\,f(y)$ for $x,y\in\CA$. This map satisfies
\begin{eqnarray}
\tilde f\big(\eps^{i_1\dots i_n}\,x^{i_1}\otimes\cdots\otimes x^{i_n}\big)&=& \eps^{i_1\dots i_n}\,f(x^{i_1})\cdots f(x^{i_n}) \nonumber\\[4pt] &=& \big[f(x^1),\dots,f(x^n)\big]_A \ = \ \tilde f\,[x^1,\dots,x^n]
\end{eqnarray}
for any $x^i\in\CA$. Thus $\tilde f$ is trivial on the ideal $I_n(\CA)$ generated by the defining relation of $U_n(\CA)$. This proves the existence of $\varphi$. Since $\CA$ generates the algebra $T(\CA)$, uniqueness follows immediately.

In order to construct a deformation quantization of a classical algebra of functions, and for consistency with our quantization axioms in which the quantum geometry is encoded in a subalgebra of $\sEnd(\CCH)$, we will deform a \emph{binary} operation. For this, let $S(\CA)$ be the symmetric polynomial algebra generated by the vector space $\CA$, i.e.\ $S(\CA)=T(\CA)/\langle x\otimes y-y\otimes x\rangle_{x,y\in\CA}$. In this case one can use the standard PBW theorem\footnote{A more general PBW theorem involving an $n$-symmetric algebra of $\CA$ can be developed along the lines of~\cite[Sec.~III.B]{Goze:2008ku}, but this is not needed here.} to set up a linear isomorphism
\begin{equation}
\xymatrix{
\psi_n\,:\, S(\CA)~ \ar[r]^{ \ \ \cong} & ~ U_n(\CA)
}
\label{PBWiso}\end{equation}
which sends a monomial $y(x^i)$ to the corresponding product of operators $\hat y(\hat x^i)$ with a suitable ordering. Given any such isomorphism, we transfer the $n$-noncommutative product on $U_n(\CA)$ to an $n$-star product on $S(\CA)$ given by
\begin{equation}
f\star_n g=\psi_n^{-1}\big(\psi_n(f)\cdot\psi_n(g)\big) \ .
\label{starprod}\end{equation}
For $k<n$, the star products $f_1\star_n\cdots\star_n f_k$, $f_i\in S(\CA)$ are generically commutative (unless some further quotient of the tensor algebra $T(\CA)$ is taken), and only for $k\geq n$ does $n$-noncommutativity generally enter. 
We regard this star product as providing a quantization of the linear Nambu-Poisson structure on the commutative algebra $S(\CA)$, associated to the $n$-Lie bracket on $\CA$ according to our generalized quantization axiom Q3\pr\ above. For $n=2$ and $\psi_2$ the standard symmetrization map, this product is equivalent to the Kontsevich star product for linear Poisson structures~\cite{Kontsevich:1997vb,Kathotia}. In the following sections we will see many examples of this quantization prescription.

\section{Berezin-Toeplitz quantization of projective spaces}

In this section we will review both Berezin and Toeplitz quantization of the complex projective spaces $\CPP^n$, as this approach will be the starting point of our ensuing discussion. The original constructions are due to Kostant and Souriau~\cite{Kostant-1970aa,Kostant:1975qe,Souriau-1970aa}, see also~\cite{IuliuLazaroiu:2008pk} for more details on the general construction and further references. The fuzzy sphere had first been discussed by Berezin in~\cite{Berezin:1974du}.
Berezin-Toeplitz quantization is a hybrid form of geometric and deformation quantization in that it uses the Hilbert space of geometric quantization together with the relaxed correspondence principle of deformation quantization. The Hilbert space is chosen as the space of holomorphic sections of a very ample line bundle over the K{\"a}hler manifold one wishes to quantize and functions turn into endomorphisms of this Hilbert space under quantization.

\subsection{Quantum line bundle on $\CPP^n$}

In the general case, we start from a complex line bundle $L$ over a K{\"a}hler manifold $\CM$ of complex dimension $n$. The line bundle is endowed with a hermitian metric $h$, and the unique connection $\nabla$ which is compatible with both the complex structure on $L$ and the metric $h$. The quantization condition states that the curvature $F$ of this connection is proportional to the K{\"a}hler form of $\CM$,
\begin{equation}\label{quantcond}
 \omega=\frac{\di}{2\pi}\,F~.
\end{equation}
In geometric quantization, this condition guarantees that the correspondence principle is satisfied. For our purposes, we merely observe that \eqref{quantcond} implies that $L$ is a positive or ample line bundle and therefore that a certain power $L^{\otimes k_0}$ of this line bundle is very ample. In the following, we will assume that $L$ is already very ample, for otherwise one can make the necessary replacements $L\rightarrow L^{\otimes k_0}$, $\omega\rightarrow k_0\,\omega$, $\nabla\rightarrow\nabla^{\otimes k_0}$, and $h\rightarrow h^{\otimes k_0}$. The line bundle $(L,h)$ is called a {\em quantum line bundle} for $(\CM,\omega)$ and $(\CM,\omega,L,h)$ is a {\em prequantized Hodge\footnote{We can choose an appropriate normalization such that $[\omega]\in H^2(\CM,\RZ)$.} manifold}. 

The hermitian metric $h$ together with the Liouville volume form $\dd\mu=\frac{\omega^n}{n!}$ on $\CM$ induces a metric on the space of smooth sections $\Gamma^\infty(\CM,L)$ given by
\begin{equation}
 (s_1|s_2):=\int_\CM\, \dd \mu~ h_x\big(s_1(x),s_2(x)\big)~,
\end{equation}
for $s_1,s_2\in \Gamma^\infty(\CM,L)$. This yields a projection from $L^2(\CM,L)$, the $L^2$-completion of the space $\Gamma^\infty(\CM,L)$, to $H^0(\CM,L)$, the space of global holomorphic sections of $L$. The inner product on $L^2(\CM,L)$ also induces a inner product on $H^0(\CM,L)$, which we denote by the same symbol. We identify $H^0(\CM,L)$ with the Hilbert space $\CCH=\CCH_L$, and by doing so we choose to work with holomorphic polarization. 

To quantize $\CM=\CPP^n$, we choose $L$ to be the holomorphic line bundle $\CO(k)$ of degree $k\in\NN$ and $\omega$ the K{\"a}hler form giving rise to the Fubini-Study metric on $\CPP^n$. For $L=\CO(k)$, the space $\CCH_k:=\CCH_L=H^0(\CM,L)$ is finite-dimensional and spanned by homogeneous polynomials of degree $k$ in the standard homogeneous coordinates $z_\alpha$, $\alpha=0,1,\dots ,n$ on $\CPP^n$. Hence
\begin{eqnarray}
 \CCH_k&:=&{\rm span}_\FC\big\{z_{\alpha_1}\cdots z_{\alpha_k}~\big|~ \alpha_i=0,1,\dots,n\big\} \nonumber\\[4pt] &=& {\rm span}_\FC\Big\{z_{0}^{p_0}\,z_1^{p_1}\cdots z_{n}^{p_n}~\Big|~ p_\alpha\in \NN_0~,~|\vec p\,|:= \mbox{$\sum\limits_{\alpha=0}^n$}\, p_\alpha=k\Big\}~.
\end{eqnarray}
For later convenience, we identify this space with the $k$-particle Hilbert space in the Fock space of $n+1$ harmonic oscillators given by
\begin{equation}
 \CCH_k\cong{\rm span}_\FC\Big\{\frac{\ah^\dagger_{\alpha_1}\cdots \ah^\dagger_{\alpha_k}}{\CN}|0\rangle\Big\}={\rm span}_\FC\Big\{\frac{(\ah^\dagger_0)^{p_0}\cdots(\ah^\dagger_n)^{p_n}}{\sqrt{p_0!\cdots p_n!}}|0\rangle =:\frac{1}{\sqrt{\vec{p}\,!}}\,|\vec{p}\, \rangle\Big\}~,
\end{equation}
where $\CN\in\FR$ is a normalization constant. The creation and annihilation operators satisfy the usual Heisenberg-Weyl algebra $[\ah_\alpha,\ah^\dagger_\beta]=\delta_{\alpha\beta}$, and $|0\rangle$ denotes the vacuum vector with $\ah_\alpha|0\rangle=0$.

\subsection{Coherent states}

Consider the total space $\mathbb{L}$ of the line bundle $L$, with projection $\pi:\mathbb{L}\rightarrow \CM$, and $\mathbb{L}_o=\mathbb{L}\backslash o$, where $o$ is the zero section. We define a function $\psi_q(s)$ which indicates how much one has to scale a section $s\in \CCH_L$ to pass through a given point $q\in\mathbb{L}_o$ via
\begin{equation}
 s\big(\pi(q)\big)=:\psi_q(s)\, q~.
\end{equation}
By Riesz's theorem, there is a unique holomorphic section $e_q$ such that
\begin{equation}
 (e_q|s)=\psi_q(s)
\end{equation}
for all sections $s\in \CCH_L$. The element $e_q$ is called the {\em Rawnsley coherent state vector}, a generalization of the {\em Perelomov coherent states} appearing from a group theoretic perspective. 
The {\em Rawnsley coherent state projector} is given by
\begin{equation}
 P_x:=\frac{|e_q)(e_q|}{(e_q|e_q)}~,~~~q\in \mathbb{L}_o~.
\end{equation}
Note that $P_x$ indeed only depends on $\pi(q)=x$. This is due to the scaling of $\psi_q$, $\psi_{c\,q}=\frac{1}{c}\,\psi_q$. 

In our quantization of $\CM=\CPP^n$ with $L=\CO(k)$, the Rawnsley
coherent states are simply the truncated Glauber vectors $|z,k\rangle$ on $\FC^{n+1}$ (cf.\ e.g.~\cite{IuliuLazaroiu:2008pk}) given by
\begin{equation}
|z\rangle=\exp\big(\bz_\alpha\, {\hat a}_\alpha^\dagger\big)|0\rangle=\sum_{\vec p} \,\frac{\bz\,^{\vec{p}}}{\sqrt{\vec{p}\,!}}\,|{\vec{p}}\,\rangle =\sum_{k=0}^\infty\, |z,k \rangle~,
\end{equation}
where
\begin{equation}
|z,k\rangle=\frac{1}{k!}\, \big({\bar
  z}_\alpha\,{\hat a}_\alpha^\dagger\big)^k|0\rangle=\sum_{|{\vec{p}}\,|=k}\,{\frac{\bz\,^{\vec{p}}}{\sqrt{{\vec{p}}\,!}}\,|{\vec{p}}\,\rangle}~.
\end{equation}
One readily verifies that the Rawnsley and Perelomov coherent states coincide in this case. The coherent state projector takes the form
\begin{equation}
 P_z=\frac{|z,k\rangle\langle z,k|}{\langle z,k|z,k\rangle}~.
\end{equation}
Useful relations for the computations which follow are $\hat a_\alpha|z,k\rangle=\bz_\alpha|z,k-1\rangle$ and $\langle z,k|z,k\rangle=\frac{1}{k!}\,|z|^{2k}$.

\subsection{Berezin quantization}

The {\em lower} or {\em covariant Berezin symbol} of an operator $\hat{f}\in\sEnd(\CCH_L)$ is defined as
\begin{equation}
 \sigma(\hat{f}\,)(x):=\tr(\hat{f}\,P_x)~.
\end{equation}
The space $\sigma(\sEnd(\CCH_L))$ is the space of quantizable functions $\Sigma\subset \CC^\infty(\CM)$. The map $\sigma$ is injective and thus we can define the Berezin quantization of a function as the inverse of $\sigma$ on $\Sigma$ given by
\begin{equation}
 f~\longmapsto~ Q(f):=\hat{f}=\sigma^{-1}(f)~,~~~f\in\Sigma~.
\end{equation}

One readily verifies the quantization axioms for $\CM=\CPP^n$. The map $Q:\Sigma\to\sEnd(\CCH_k)$ is linear, and the constant function is indeed mapped to the identity operator since from the form of the coherent state projector we find
\begin{equation}
 Q\Big(\frac{z_{\alpha_1}\cdots z_{\alpha_k}\, \bz_{\beta_1}\cdots \bz_{\beta_k}}{|z|^{2k}}\Big)=\frac{1}{k!}\, \ah^\dagger_{\alpha_1}\cdots\ah^\dagger_{\alpha_k}|0\rangle\langle 0|\ah_{\beta_1}\cdots\ah_{\beta_k}
\end{equation}
 with $|z|^2:=\bz_\alpha\, z_\alpha$, so that in particular $Q(1)=\unit_{\CCH_k}$. To check the third quantization axiom, it is convenient to employ a ``star product\footnote{This product, sometimes called the coherent state star product, is not a formal star product.}'' on $\CPP^n$. As usual, the star product is induced by pulling back the operator product onto the set of quantizable functions to get
\begin{equation}
 f*g:=\sigma(\hat{f}\,\hat{g})~,~~~f,g\in\Sigma~.
\end{equation}
To obtain a particularly nice form, one needs an embedding $\CPP^n\embd \FR^{(n+1)^2-1}$ given by the Jordan-Schwinger transformation
\begin{equation}\label{JordanSchwingerTrafo}
 x^M=\frac{\bz_\alpha \,\lambda_{\alpha\beta}^M\, z_\beta}{|z|^2}~,
 \qquad M=1,\dots,(n+1)^2-1 \ ,
\end{equation}
where $\lambda^M_{\alpha\beta}$ are the Gell-Mann matrices of the isometry group $\sSU(n+1)$ of $\CPP^n$. In terms of the coordinates $x^M$, one can write this star product as~\cite{Balachandran:2001dd}
\begin{equation}\label{cpnstarproduct}
 (f*g)(x)=\sum_{l=0}^k\, \frac{(k-l)!}{k!\, l!}\, \big(\dpar_{M_1}\cdots \dpar_{M_l}f(x)\big)\, K^{M_1N_1}\cdots K^{M_lN_l}\,\big(\dpar_{N_1}\cdots \dpar_{N_l}g(x) \big)~,
\end{equation}
where $\dpar_M:=\der{x^M}$ and 
\begin{equation}
 K^{MN}=\frac{1}{n+1}\, \delta_{MN}+\frac{1}{\sqrt{2}}\, \big(d^{MN}{}_K+\di\, f^{MN}{}_K\big)\, x^K-x^M\, x^N~.
\end{equation}
Here $d^{MN}{}_K$ and $f^{MN}{}_K$ are the symmetric tensor and structure constants of $\sSU(n+1)$. Note that \eqref{cpnstarproduct} forms an expansion in terms of $\hbar=\frac{1}{k}$ for $k$ large. One can derive that the symplectic form which gives rise to the Fubini-Study metric on $\CPP^n$ in the coordinates $x^M$ is given by $2\,\di\, K^{[MN]}$~\cite{Balachandran:2001dd}. The correspondence principle therefore reads as
\begin{equation}
\lim_{k\rightarrow \infty}\, \big\|\,\di\, k\, (f*g-g*f)-2\, \di\,
K^{[MN]}\, (\dpar_M f)\, (\dpar_N g)\big\|_{L^2}=0~,
\end{equation}
which one readily verifies using \eqref{cpnstarproduct}.

Let us examine the case of $\CPP^1\cong S^2$ in some more detail. With the choice $L=\CO(k)$, $\Sigma$ corresponds to the set of spherical harmonics $Y_{\ell m}$ with angular momentum $\ell\leq k$. The Poisson bracket is
\begin{equation}
 \{x^\mu,x^\nu\}=R\,\eps^{\mu\nu\kappa}\, x^\kappa~,
\end{equation}
where $R$ is the radius of the sphere $S^2$. The quantization axiom Q3 then implies that the quantizations $\xh^\mu$ of the coordinates $x^\mu$ satisfy the Lie algebra
\begin{equation}\label{S2relation}
 [\xh^\mu,\xh^\nu]= -\di\, \hbar\,R\, \eps^{\mu\nu\kappa}\, \xh^\kappa~.
\end{equation}
The deformation parameter $\hbar$ here is not continuous. To compute it, we again use the Jordan-Schwinger transformation \eqref{JordanSchwingerTrafo},
\begin{equation}
 x^\mu:=\frac{R}{|z|^2}\,\bz_\alpha\, \sigma^\mu_{\alpha\beta}\, z_\beta~,
\end{equation}
where $x^\mu$ are coordinates
on $S^2\,\embd\, \FR^3$ and $z_\alpha$ are homogeneous coordinates on
the projective line $\CPP^1$, while $\sigma^\mu$, $\mu=1,2,3$, is the
standard basis of $2\times2$ Pauli spin matrices for $\asu(2)$, see Appendix~\ref{Cliffalg} One can easily work out the quantization of the coordinate functions to be
\begin{equation}
 x^\mu~\longmapsto~
 \xh^\mu=\frac{R}{k!}\,\sigma^\mu_{\alpha\beta}\,\hat a^\dagger_\alpha\, \hat a^\dagger_{\rho_1}\cdots \hat a^\dagger_{\rho_{k-1}}|0\rangle\langle 0|\hat a_\beta\, \hat a_{\rho_1}\cdots \hat a_{\rho_{k-1}}=:\frac{R}{k!}\, \sigma^\mu_{\alpha\beta}\,\st{\alpha}{\beta}~.
\end{equation}
Working through the details, one obtains
\begin{equation}
 \hbar=\frac{2}{k}~.
\end{equation}
The classical limit is obtained for $k\rightarrow\infty$, and \eqref{S2relation} suggests that in this limit the algebra of coordinate functions (and thus the whole algebra of functions) becomes indeed commutative. For details on how to construct a deformation quantization as well as a formal star product out of algebras of covariant Berezin symbols in this case, see~\cite{IuliuLazaroiu:2008pk} and references therein.

\subsection{Toeplitz quantization}

In Toeplitz quantization (see e.g.~\cite{Bordemann:1993zv}), the operator $T_L(f)$ corresponding to a function $f$ acts on an element $s$ of the Hilbert space $\CCH_L$ by multiplying the corresponding section $s$ and subsequent projection back to holomorphic sections via the inner product $(-|-)$. Hence
\begin{equation}
 T_L(f)(s):=\Pi(f\,s)~,~~~f\in \CC^\infty(\CM)~,~s\in \CCH_L~.
\end{equation}
The appropriate projector is evidently the coherent state projector $P_x$ and we thus arrive at
\begin{equation}
 T_L(f)=\int_\CM\, \dd \mu~f(x)\, P_x~.
\end{equation}
The Toeplitz quantization map is the adjoint of the Berezin quantization map with respect to the Hilbert-Schmidt norm and the $L^2$-measure induced by the Liouville volume form~\cite{Schlichenmaier-1999aa}. The ordering prescriptions resulting from Berezin and Toeplitz quantizations of $\CM=\CPP^n$ correspond to Wick and anti-Wick ordering, respectively, cf.~\cite{IuliuLazaroiu:2008pk}.

Toeplitz quantization is of interest for various reasons. First, it converges towards geometric quantization as shown in~\cite{Tuynman:1987jc}. Second, strict convergence theorems can be deduced, and in particular for $\CM=\CPP^n$ one has~\cite{Bordemann:1993zv}
\begin{equation}
 \lim_{k\rightarrow \infty}\, \Big\|\,\di\, k\, \big[T_{\CO(k)}(f),T_{\CO(k)}(g)\big]-T_{\CO(k)}\big({\{f,g\}} \big)\,\Big\|_{HS}=0~.
\label{CPnlim}\end{equation}
In Toeplitz quantization, the quantization map is not bijective and therefore to apply the correspondence principle in the form Q3\pr, one has to restrict to the set of quantizable functions first as done in the case of Berezin quantization.

\subsection{Scalar field theory on quantum projective spaces}

To write down the action of scalar field theory on Berezin-quantized $\CPP^n$, we need two additional structures: a Laplace operator and an integration.
To obtain the latter, consider two sections $s,t\in \CCH_L$. At $x=\pi(q)$, we have
\begin{equation}
 s(x)=\psi_q(s)\, q \ \eand \ t(x)=\psi_q(t)\, q~,
\end{equation}
where $\psi_q$ is the scaling function introduced above and $q\in\mathbb{L}_o$. The {\em Rawnsley $\eps$-function} is defined as~\cite{Rawnsley:1976gb}
\begin{equation}
 \eps\big(\pi(q)\big):=h_{\pi(q)}(q,q)~(e_q|e_q)~,
\end{equation}
and it allows us to write down the relation
\begin{equation}
\begin{aligned}
 (s|t)&=\int_\CM\, \dd \mu~h_x\big(s(x),t(x)\big)\\[4pt] &=\int_\CM\, \dd \mu~ \overline{\psi_q(s)}\, \psi_q(t)~h_{\pi(q)}(q,q)\\[4pt] &=\int_\CM\, \dd \mu ~(s|e_q)\,(e_q|t)~h_{\pi(q)}(q,q)=\int_\CM\, \dd \mu~\eps(x)~(s|P_x|t)~.
\end{aligned}
\end{equation}
Considering arbitrary sections $s,t\in \CCH_L$, we obtain the overcompleteness relation
\begin{equation}
 \int_\CM\, \dd \mu~\eps(x)~P_x=\unit_{\CCH_k}~,
\end{equation}
and hence
\begin{equation}
 \int_\CM\, \dd \mu~\eps(x)~f(x)=\tr\big(\sigma^{-1} (f) \big)~.
\end{equation}

There are two ways of carrying over a linear operator $\CD$ from the algebra of smooth functions $\CC^\infty(\CM)$ to $\sEnd(\CCH_L)$~\cite{IuliuLazaroiu:2008pk}, the {\em Berezin-push} $\CD^B$ and the {\em Berezin-Toeplitz lift} $\CD_\diamond$. They are defined according to
\begin{equation}
 \CD^B:=Q\circ \CD\circ \sigma \ \eand \ \CD_\diamond:=T_L \circ \CD\circ \sigma~.
\label{pushlift}\end{equation}
While the former guarantees that the identity operator on $\CC^\infty(\CM)$ is mapped to the identity operator on $\sEnd(\CCH_L)$, the latter ensures that hermitian operators with respect to the canonical $L^2$-measure on $\CC^\infty(\CM)$ are mapped to hermitian operators with respect to the Hilbert-Schmidt norm.

For $\CM=\CPP^n$ and $\omega$ the K{\"a}hler form giving rise to the Fubini-Study metric, $\eps(x)=\frac{(n+k)!}{{\rm vol}(\CPP^n)\,n!\, k!}$ is a constant, cf.~\cite{IuliuLazaroiu:2008pk}. Moreover, both definitions in (\ref{pushlift}) agree up to a constant~\cite{IuliuLazaroiu:2008pk}. The Hilbert space $\CCH_k$ carries an irreducible representation of the isometry group $\sSU(n+1)$ of $\CPP^n$. The Berezin-Toeplitz lift (as well as the Berezin-push) of the Laplace operator corresponds to the quadratic Casimir operator of $\sSU(n+1)$ in this representation.

This completes the construction of the necessary ingredients for writing down an action functional for a scalar field theory on fuzzy $\CPP^n$. Analyzing the corresponding partition function is facilitated by the fact that the configuration space is finite-dimensional, and thus the functional integral is well-defined. For this reason, fuzzy $\CPP^n$ can be used as a regulator for quantum field theories.

\section{Quantization of spheres\label{Qspheres}}

In this section, we will provide an extension of Berezin-Toeplitz quantization to spheres. We shall also examine in detail the $n$-Lie algebra structure on the arising operator algebra and compare these quantizations to previous versions of fuzzy spheres in higher dimensions.

\subsection{Hyperspherical harmonics}

Consider the space $\FR^{d+1}$ with its usual cartesian coordinates
$x^\mu$, $\mu=1,\dots,d+1$. Let $S^d$ be the sphere of radius $R$
embedded in this space as the quadric $x^\mu \, x^\mu=R^2$. The hyperspherical harmonics $Y_{\ell\mbf m}$ spanning the algebra of smooth functions $\CC^\infty(S^d)$ correspond to polynomials which are of degree $\ell$ in the coordinates $x^\mu$ after imposing the equation $x^\mu\, x^\mu=R^2$.

There is an embedding of even-dimensional spheres $S^d$ into $\CPP^r$, with $r+1:=2^{\lfloor\frac{d+1}{2}\rfloor}$ the dimension of the spinor representation of $\sSO(d+1)$. For this, consider the generators $\gamma^\mu$, $\mu=1,\dots,d+1$, of the Clifford algebra\footnote{A construction of the explicit matrix representation of the Clifford algebras yielding spinor representations is given in Appendix~\ref{Cliffalg}} $Cl(\FR^{d+1})$ satisfying $\{\gamma^\mu,\gamma^\nu\}=2\delta^{\mu\nu}$. If $d$ is even, the spinor representation of $\sSO(d+1)$ is irreducible. The readily verified relation\footnote{Here and in the following, $\odot$ denotes the normalized symmetric tensor product.}
\begin{equation}
 [\gamma^{\mu\nu}\odot\unit_{r+1},\gamma^\rho\odot\gamma^\rho]=0 \ ,
\end{equation}
where $\gamma^{\mu\nu}:=\frac12\,[\gamma^\mu,\gamma^\nu]$, together
with Schur's lemma implies 
$\gamma^\rho\odot\gamma^\rho= c\,\unit_{r+1}\odot\unit_{r+1}$,
$c\in\FC$, for even $d$. Using the generators $\gamma^\mu_{\alpha\beta}$ of the
Clifford algebra constructed in Appendix~\ref{Cliffalg} yields $c=1$,
so $\gamma^\rho\odot\gamma^\rho=\unit_{r+1}\odot\unit_{r+1}$. Therefore, the embedding relation $x^\mu\, x^\mu=R^2$ is satisfied for 
\begin{equation}\label{embedding}
x^\mu:=\frac{R}{|z|^2}\, \bz_\alpha \,\gamma^\mu_{\alpha\beta}\, z_\beta~,
\end{equation}
which generalizes the usual Jordan-Schwinger transformation.
The space of hyperspherical harmonics $Y_{\ell \mbf m}$ with $\ell\leq k$ is thus spanned by the functions
\begin{equation}
 \gamma^{\mu_1}_{\alpha_1\beta_1}\cdots\gamma^{\mu_j}_{\alpha_j\beta_j}\, \delta_{\alpha_{j+1}\beta_{j+1}}\cdots\delta_{\alpha_k\beta_k}\, \bz_{\alpha_1}\cdots\bz_{\alpha_k}\, z_{\beta_1}\cdots z_{\beta_k}~,
\end{equation}
and they transform in the product of two totally symmetric tensor product representations
\begin{equation}
\underbrace{\overline{{\tyng(6)}}}_{\mbox{\scriptsize{$k$~boxes}}} ~ \otimes ~ 
\underbrace{{\tyng(6)}}_{\mbox{\scriptsize{$k$~boxes}}}
\end{equation}
of $\sSO(d+1)\simeq \sSpin(d+1)$.

Our embedding $S^d \, \embd \, \CPP^r$ induces an injection $\rho:\CC^\infty(S^d) \, \embd \, \CC^\infty(\CPP^r)$. Polynomials in the coordinates $x^\mu$ restricted to $S^d$ form a dense subset in $\CC^\infty(S^d)$ and they are turned into global functions on $\CPP^r$ via the substitution \eqref{embedding}. Moreover, the Fubini-Study metric on $\CPP^r$ induces the standard round metric on $S^d$, with volume form $\dd\mu_{S^d}$, which is easily seen as the embedding is manifestly $\sSO(d+1)$-invariant. This implies in particular that for a function $f\in \CC^\infty(\CM)$, one has
\begin{equation}\label{ReduceIntegral}
 \int_{\CPP^r}\, \dd \mu~ \rho(f)= {\rm vol}~ \int_{S^d}\, \dd \mu_{S^d}~ f~,
\end{equation}
where ${\rm vol}$ is a constant volume factor. Therefore, the $L^2$-inner product on $\CPP^r$ with respect to the Fubini-Study metric is naturally compatible with the $L^2$-inner product on $S^d$ with respect to the round metric.

Equivalently, we can consider the tensor algebra $\aT$ generated by
the $x^\mu$ and factor by the ideal $\aI$ generated by
$R^2\,1-x^\mu\otimes x^\mu$ as well as $x^\mu\otimes
x^\nu-x^\nu\otimes x^\mu$. Note that $1$ here is the unit of $\FC$
and thus we have $1\otimes x^\mu=x^\mu$. The algebra of smooth
functions on $S^d$ is indeed isomorphic to a suitable completion of
the polynomial quotient algebra $\aT/\aI$.

We will obtain odd-dimensional spheres as a reduction of even-dimensional spheres. We reduce $S^{2d}$ to $S^{2d-1}$ by putting $x^{2d+1}=0$. Let us introduce $s:=\frac{r+1}{2}$. Using the inductive construction of the Clifford algebra given in Appendix~\ref{Cliffalg}, we have $\gamma^{2d+1}=\di^d\,\unit_s\otimes\sigma^3$, where the gamma-matrices act on $\FC^{r+1}=\FC^{2s}$. In complex coordinates, the condition $x^{2d+1}=0$ thus implies
\begin{equation}
 \sum_{\alpha=0}^{s-1}\,\bz_{\alpha}\, z_\alpha-\sum_{\alpha=s}^{2s}\, \bz_{\alpha}\, z_\alpha=0~.
\end{equation}
This condition reduces the space $\CPP^r$, into which we embedded $S^{2d}$, to $\CPP^{s-1}\times\CPP^{s-1}$. In particular, this reduces the embedding $S^4\,\embd \,\CPP^3$ to $S^3\,\embd\, \CPP^1\times\CPP^1$. 

We can go one step further and reduce $S^{2d-1}$ to $S^{2d-2}$ by putting $x^{2d}=0$. In the inductive construction, we have $\gamma^{2d}=\unit_s\otimes \sigma^1$, which yields the condition
\begin{equation}
 \sum_{\alpha=0}^{s-1}\, (\bz_{\alpha}\,z_{\alpha+s} +\bz_{\alpha+s}\, z_{\alpha})=0~.
\end{equation}
This equation is solved by putting $z_{\alpha+s}=\di\, z_\alpha$,
which reduces $\CPP^{s-1}\times \CPP^{s-1}$ to the diagonal subspace $\CPP^{s-1}$.
It follows from both the reduction as well as the fact that the embedding respects the isometries that \eqref{ReduceIntegral} also holds for odd-dimensional spheres.

\subsection{Berezin quantization of even-dimensional spheres}

Even-dimensional spheres $S^d$ are rather straightforward
to quantize, and we therefore start with them. Our goal is to
construct a Hilbert space $\CCH_k$ together with a quantization map
$x^\mu\mapsto\hat x^\mu$ taking functions on $S^d$ to endomorphisms of
$\CCH_k$ such that $\xh^\mu\, \xh^\mu=R_F^2\,\unit_{\CCH_k}$, where the ``fuzzy radius'' $R_F$ will be identified below. We also want to construct the bracket of a $d$-Lie algebra, such that ideally it satisfies the generalized quantization axiom Q3\pr. For the spheres, this implies that we are looking for a quantization map $x^\mu\mapsto \xh^\mu$ together with a $d$-Lie bracket satisfying
\begin{equation}
 [\xh^{\mu_1},\dots,\xh^{\mu_d}]=-\di\, \hbar(k)\, R^{d-1}\,
 \eps^{\mu_1\dots\mu_d\mu_{d+1}}\, \xh^{\mu_{d+1}}~.
\label{dLieSd}\end{equation}

We return to the embedding of $S^d$ into $\CPP^r$ and use the Hilbert
space $\CCH_k$ of Berezin-quantized $\CPP^r$ with quantum line bundle
$L=\CO(k)$. Thus $\CCH_k$ is identified as the $k$-particle subspace
of the Fock space of $r+1$ harmonic oscillators, with
\begin{equation}
 \ah_{\alpha_1}^\dagger\cdots \ah_{\alpha_k}^\dagger|0\rangle\in \CCH_k~,~~~[\ah_\alpha,\ah^\dagger_\beta]=\delta_{\alpha\beta}~,~~~\ah_\alpha|0\rangle=0~.
\end{equation}
We define the lower Berezin symbol $\sigma_R(\hat{f}\,)$ of an
operator $\hat{f}\in\sEnd(\CCH_k)$ by the $L^2$-projection of the
lower Berezin symbol $\sigma(\hat{f}\, )\in \Sigma\subset \CC^\infty(\CPP^r)$ onto $\Sigma_R \subset \CC^\infty(S^d)$. Explicitly, this amounts to introducing the restricted coherent state projector
\begin{equation}
\begin{aligned}
 P^R_x:=&\sum_{m=0}^k\, x^{\mu_1}\cdots
 x^{\mu_m}\, k!\,\left(\frac{2}{R}\right)^m\,
 \gamma^{\mu_1}_{\alpha_1\beta_1}\cdots
 \gamma^{\mu_m}_{\alpha_m\beta_m}\\ & \qquad\qquad \times \,
 \ah^\dagger_{\alpha_1}\cdots \ah^\dagger_{\alpha_m}\,
 \ah^\dagger_{\rho_{1}}\cdots \ah^\dagger_{\rho_{k-m}}|0\rangle\langle
 0|\ah_{\beta_1}\cdots \ah_{\beta_m}\, \ah_{\rho_{1}}\cdots
 \ah_{\rho_{k-m}}\\[4pt] 
=:& \sum_{m=0}^k \, x^{\mu_1}\cdots x^{\mu_m}\, k!\,
\left(\frac{2}{R}\right)^m\, \gamma^{\mu_1}_{\alpha_1\beta_1}\cdots
\gamma^{\mu_m}_{\alpha_m\beta_m}\, \st{{\alpha_1}\dots {\alpha_m}}{{\beta_1}\dots {\beta_m}}~,
\end{aligned}
\end{equation}
and with eq.~\eqref{gammadecompose} of Appendix~\ref{Cliffalg} we
conclude that indeed $P^R_x\,P^R_x=P^R_x$. The coordinates $x^\mu$ can
be substituted again by \eqref{embedding} to obtain an expression for
$P_x^R$ in terms of homogeneous coordinates on $\CPP^r$. The restriction of the lower Berezin symbol now reads
\begin{equation}
 \sigma_R(\hat{f}\, )(x) :=\tr(P_x^R\, \hat{f}\,)~.
\end{equation}

The map $\sigma_R:\sEnd(\CCH_k)\rightarrow \Sigma_R$ is no longer injective due to the projection involved from $\Sigma\subset \CC^\infty(\CPP^r)$ to $\Sigma_R$. However, since $\Sigma_R\subset \Sigma$, we can use the inverse of the unrestricted Berezin symbol $\sigma$ to define a quantization map
\begin{equation}
 Q\,:\, \Sigma_R~\longrightarrow~ \sEnd(\CCH_k) \ , \qquad f~\longmapsto~ \sigma^{-1}(f)~.
\end{equation}
For the coordinate functions, this quantization yields
\begin{equation}
 x^\mu~ \longmapsto~ \xh^\mu:=Q(x^\mu)=\frac{R}{k!}\,
 \gamma^\mu_{\alpha\beta} \, \st{\alpha}{\beta}~.
\end{equation}

The operators $\xh^\mu$ indeed generate all of $\sEnd(\CCH_k)$. For
this, we first note that totally antisymmetric products of $d-1$ of
the operators $\xh^\mu$ span the space of all operators of the form $\st{\alpha_1}{\beta_1}$. A product of two such antisymmetric products decomposes into operators of the form $\st{\alpha_1\alpha_2}{\beta_1\beta_2}$ and $\st{\alpha_1}{\beta_1}$. In this way, we can inductively construct all of $\sEnd(\CCH_k)$ by noncommutative polynomials in the operators $\xh^\mu$ of maximal degree $k\, (d-1)$. This implies in particular that the noncommutative polynomials of degree $k\, (d-1)$ form an algebra. This agrees with the known result for the fuzzy sphere, where the algebra $\sEnd(\CCH_k)$ consists of noncommutative polynomials of degree $k$. 

This quantization clearly satisfies the quantization axioms Q1, Q2, and Q4\pr, as these properties trivially survive the projection. We will come back to the $d$-Lie algebra structure and the correspondence principle Q3\pr\ shortly.

\subsection{Toeplitz quantization of spheres}

Recall that the embedding \eqref{embedding} induces an injection $\rho:\CC^\infty(S^d)\, \embd\, \CC^\infty(\CPP^r)$. We can therefore identify the Toeplitz quantization of a function $f\in\CC^\infty(S^d)$ with the Toeplitz quantization of $\rho(f)\in\CC^\infty(\CPP^r)$. This means, in particular, that the convergence theorems of~\cite{Bordemann:1993zv} hold on $S^d$ as well. For this, recall that for $\CM=\CPP^r$ we have
\begin{equation}\label{ToeplitzEstimates}
\lim_{k\rightarrow \infty} \,\big\| T_{\CO(k)}(f)\big\|_{HS}=\|f\|_{L^2}
\end{equation}
together with (\ref{CPnlim}). On $S^d$, we consider the Poisson
structure which is obtained via the pull-back of the symplectic form
$\omega$ along the embedding $S^d\, \embd \, \CPP^r$. It follows that
the Poisson algebra thus obtained on $S^d$ is embedded in the Poisson
algebra on $\CPP^r$, and the estimates (\ref{CPnlim}) and \eqref{ToeplitzEstimates} for $S^d$ are just restrictions of the corresponding estimates on $\CPP^r$.

\subsection{$d$-Lie algebra structure}

As discussed in Section~\ref{NPquant}, we will use the $d$-Lie bracket constructed out of a lift of the truncation of the Nambu-Poisson structure on $\Sigma_R$. For this, note that $\Sigma_R$ consists of polynomials in the $x^\mu$ of maximal degree $k$, and that the components of the Nambu-Poisson tensor are homogeneous polynomials of degree 1. We can therefore endow $\Sigma_R$ with the truncated Nambu-Poisson bracket $\{-,\ldots,-\}_k$. Furthermore, we lift this bracket to $\sEnd(\CCH)$ as described in Section \ref{NPquant} The resulting $d$-Lie bracket satisfies the correspondence principle by definition. Note that it vanishes on operators $\Ah\in\sEnd(\CCH)$ with vanishing Berezin symbol $\sigma_R(\Ah)$. 

Let us now examine how this bracket is related to the totally antisymmetric operator product \eqref{antisymop}. First, note that
\begin{equation}
 {}[\xh^1,\ldots,\xh^d]=-\di\hbar \xh^{d+1}~.
\end{equation}
The antisymmetric product of two operators is given by
\begin{equation}
\begin{aligned}\label{operatorproduct}
 \hat x^\mu\,\hat x^\nu & = \left(\frac{R}{k!}\right)^2\,
 \big(\gamma^\mu_{\alpha\beta}\,\st{\alpha}{\beta}\big)\,
 \big(\gamma^\nu_{\gamma\delta}\, \st{\gamma}{\delta}\big)\\[4pt] 
&=\frac{R^2}{k!\,k}\, (\gamma^\mu\,\gamma^\nu)_{\alpha\beta}\,
\st{\alpha}{\beta}+\frac{R^2\,(k-1)}{k!\,k}\,
\gamma^\mu_{\alpha_1\beta_1}\, \gamma^\nu_{\alpha_2\beta_2}\, \st{\alpha_1\alpha_2}{\beta_1\beta_2}~.
\end{aligned}
\end{equation}
Due to $\sSO(d+1)$-invariance of the construction, we can focus on the expression $[\xh^1,\ldots,\xh^d]$. Using \eqref{operatorproduct} we compute
\begin{equation}
 \sum_{\mu_i=1}^d\, \eps^{\mu_1\ldots\mu_d}\, \xh^{\mu_1}\cdots\xh^{\mu_d}
=\sum_{\mu_i=1}^d\, \eps^{\mu_1\ldots\mu_d}\,\xh^{\mu_1\mu_2}\cdots\xh^{\mu_{d-1}\mu_d}~,
\end{equation}
where we introduced $\xh^{\mu\nu}:=\tfrac{1}{2}\, [\xh^\mu,\xh^\nu]=\frac{R^2}{k!\,k}\, (\gamma^{\mu\nu})_{\alpha\beta}\,
\st{\alpha}{\beta}$. From \eqref{operatorproduct}, we notice that for
large $k$ the dominant contribution to the above $d$-bracket
is given by
\begin{equation}
\begin{aligned}
 &\sum_{\mu_i=1}^d\, \eps^{\mu_1\ldots\mu_d}\,
 \left(\frac{R^2}{k!\,k}\right)^{\frac{d}{2}}\, \big((k-1)\,(k-1)!
 \big)^{\frac{d}{2}-1}\\&\hspace{3cm}\times\, (\gamma^{\mu_1\mu_2})_{\alpha_1\alpha_2}\cdots (\gamma^{\mu_{d-1}\mu_{d}})_{\alpha_{d-1}\alpha_{d}}\, \st{\alpha_1\alpha_3\ldots\alpha_{d-1}}{\alpha_2\alpha_4\ldots\alpha_d}~.
\end{aligned}
\end{equation}
Thus we have to study the symmetric tensor product
\begin{equation}
 \sum_{\mu_i=1}^d\, \eps^{\mu_1\ldots\mu_d}\, \gamma^{\mu_1\mu_2}\odot\cdots\odot\gamma^{\mu_{d-1}\mu_d}~,
\end{equation}
and the desired outcome would be proportional to $\gamma^{d+1}\odot
\unit_{r+1}$ (and hence the full result to~$\xh^{d+1}$).

Together with the formulas given in Appendix~\ref{appTPformulas}, one can evaluate this product for various $d$. For example, for $d=4$ we have
\begin{equation}
 \sum_{\mu_i=1}^4\, \eps^{\mu_1\mu_2\mu_3\mu_4}\,
 \gamma^{\mu_1\mu_2}\odot\gamma^{\mu_3\mu_4}=-\sum_{\mu_i=1}^4\,
 \gamma^5\, \gamma^{\mu_3\mu_4}\odot\gamma^{\mu_3\mu_4}=4\gamma^5\odot\unit_4~.
\end{equation}
Including all orders in $k$ we find
\begin{equation}
{}[\xh^{\mu_1},\xh^{\mu_2},\xh^{\mu_3},\xh^{\mu_4}]=8R^3\,\frac{k+2}{k^3}\,
\eps^{\mu_1\mu_2\mu_3\mu_4\mu_5}\, \xh^{\mu_5}~.
\end{equation}
This agreement between the totally antisymmetric operator product and the $d$-Lie bracket on $\sEnd(\CCH)$ breaks down, however, for polynomials of higher degree: While the latter $d$-ary product satisfies the fundamental identity for arbitrary operators, the former does not. Also, performing the same calculation for $d=6$, one concludes that both $d$-ary products do not agree here even for linear polynomials. The same feature is expected to hold for higher $d$. Summarizing, the $d$-Lie bracket agrees with the totally antisymmetric operator product for linear polynomials and $d\leq 4$.

\subsection{Commutative limit}

A nice feature of the rather explicit quantization prescription given above is that the
commutative limit is intuitively very clear. Consider again the
product \eqref{operatorproduct}. While the first term receives
contributions from both symmetric and antisymmetric parts in $\mu$ and
$\nu$, the second term is symmetric. The first term is also relatively
suppressed by a factor of $k-1$. It therefore vanishes in the limit
$k\rightarrow\infty$, rendering the coordinate algebra
commutative. Analogously, one can show that the nonassociativity for
odd-dimensional spheres (see below) vanishes in the limit, cf.~\cite{Berman:2006eu}.

The radius of the fuzzy spheres is defined through
$\xh^\mu\,\xh^\mu=R^2_F\, \unit_{\CCH_k}$. One readily computes
\begin{equation}
 \xh^\mu\,\xh^\mu=R^2\,\Big(1+\frac{d}{k}\Big)\,
 \unit_{\CCH_k}~,~~~\unit_{\CCH_k}=\frac{1}{k!} \, \unitH~.
\end{equation}
In the limit $k\rightarrow \infty$, the fuzzy radius
$R_F=\sqrt{1+\frac{d}{k}}\, R$ approaches the classical radius of $S^d$.

\subsection{Quantized isometries}

We now examine how the $\sSO(d+1)$ isometries of the sphere translate to quantum level. Recall first that on $\Sigma_R$, the rotations act according to 
\begin{equation}
 M^{\mu\nu} \acton f:= x^\mu\dpar^\nu f-x^\nu\dpar^\mu f\sim\bz^\alpha\, \gamma^{\mu\nu}_{\alpha\beta}\,
 \der{\bz^\beta}f-z^\alpha \,\gamma^{\mu\nu}_{\alpha\beta}\, \der{z^\beta}f~.
\end{equation}
This action is contained in the associated Lie algebra of the $d$-Lie algebra $\Sigma_R$. Note that the lift of this structure to $\sEnd(\CCH)$ produces the correct action of $\sSO(d+1)$ only for operators $\Ah$, for which $(\sigma^{-1}\circ\sigma)(\Ah)=\Ah$. Operators $\Ah$ for which $\sigma(\Ah)=0$ are obviously left invariant under the action of $\frg_{\sEnd(\CCH)}$.

Note also that the associated Lie algebra of the $d$-Lie algebra $\Sigma_R$ contains a subset of the diffeomorphisms, as well, which is similarly translated appropriately only to some operators in $\sEnd(\CCH)$.

\subsection{Odd-dimensional spheres}

The quantization of odd-dimensional spheres is slightly more
subtle. We want to obtain the odd spheres $S^{2d-1}$ from the even
spheres $S^{2d}$ by some kind of reduction process.  A naive approach
would be to
translate the constraint $x^{2d+1}=0$ to the operator equation
$\xh^{2d+1}|\mu\rangle=0$ for all $|\mu\rangle\in\CCH_k$. This
approach does not work,\footnote{Nor does the slight generalization
  $\xh^{2d+1}\, \xh^{2d+1}|\mu\rangle=0$.} as the condition is not invariant under the action of operators corresponding to other coordinates. The underlying reason is that the Hilbert space $\CCH_k$ corresponds to a subring of the homogeneous coordinate ring of $\CPP^r$, and imposing operator conditions on the Hilbert space corresponds therefore to factoring by a holomorphic ideal, cf.~\cite{Saemann:2006gf}. The condition $x^{2d+1}=0$, however, is not holomorphic.

The main problem here is that although we still have
$[\gamma^{\mu\nu}\odot\unit_{r+1},\gamma^\rho\odot\gamma^\rho]=0$, Schur's
lemma does not apply as the representation is reducible. It is
therefore necessary to restrict to a maximal set of irreducible
representations on which
$\xh^\mu\,\xh^\mu=R_F^2\,\unit_{\CCH_k}$. The
construction~\cite{Guralnik:2000pb,Ramgoolam:2001zx} is rather
technical, and so we just comment on the interpretation in terms of
oscillators.

For simplicity, consider $S^3\, \embd\, \CPP^1\times \CPP^1\subset \CPP^3$. We split the annihilation and creation operators of the harmonic oscillators appearing in the quantization of $\CPP^3$, $\ah_\alpha,\ah^\dagger_\alpha$, $\alpha=0,1,2 ,3$, into two groups of harmonic oscillators appearing in the quantization of $\CPP^1\times \CPP^1$, $\hat{b}_\beta,\hat{b}^\dagger_\beta$ and $\hat{c}_\beta,\hat{c}^\dagger_\beta$, $\beta=0,1$. The reduced Hilbert space is spanned by the two classes of vectors
\begin{equation}
 \hat{b}^\dagger_{\beta_1}\cdots \hat{b}^\dagger_{\beta_{s-1}}\,
 \hat{c}^\dagger_{\beta_{s}}\cdots
 \hat{c}^\dagger_{\beta_k}|0\rangle\in
 \CV_{k,s-1}~,~~~\hat{b}^\dagger_{\beta_1}\cdots
 \hat{b}^\dagger_{\beta_s}\, \hat{c}^\dagger_{\beta_{s+1}}\cdots \hat{c}^\dagger_{\beta_k}|0\rangle\in\CV_{k,s}~,
\end{equation}
where $s=\frac{k+1}{2}$ and $k$ is restricted to odd values. The
operator product is always followed by a projection back onto this
Hilbert space, which renders it nonassociative. On the irreducible
representations $\CV_{k,s}$ and $\CV_{k,s-1}$ of $\sSpin(4)$, the
operator product $\xh^\mu\, \xh^\mu$ is indeed proportional to
the identity operator. For this, recall that
\begin{equation}
\CV_k:=\bigoplus_{s=0}^k\, \CV_{k,s}
\end{equation}
is an irreducible representation of $\sSpin(5)$. Since
$(\gamma^\mu)^2\propto \unit_4$, it suffices to examine the eigenvalues
of the operator $\CO:=\gamma^5\odot\gamma^5\odot\unit_4\odot\dots
\odot\unit_4$, where $\gamma^5=-\gamma^1\,\gamma^2\, \gamma^3\, \gamma^4$. In Appendix~\ref{gammach}, it is shown that the eigenvalues of $\CO$ in the representations $\CV_{k,s}$ and $\CV_{k,k-s}$ are identical.
Moreover, on $S^3$, the totally antisymmetric operator product which agrees with the $3$-Lie bracket at linear level should actually be modified to read as
\begin{equation}
 [\xh^\mu,\xh^\nu,\xh^\kappa]:=-[\xh^\mu,\xh^\nu,\xh^\kappa,\xh^5]=
 \di\, \hbar(k)\, R^2\, \eps^{\mu\nu\kappa\lambda}\, \xh^\lambda~,
\end{equation}
which has been suggested in~\cite{Basu:2004ed}. Because of these technicalities, we have focused our discussion on even-dimensional spheres with the extensions to odd-dimensional spheres being obvious.

\subsection{Comparison to other fuzzy spheres}

Let us now put our quantization prescription into the context of previous constructions of fuzzy spheres. First, the idea of embedding spheres into complex projective space has been used previously to construct fuzzy spheres. In particular, the fuzzy 4-sphere has been constructed from the fact that $\CPP^3$ is a sphere bundle over $S^4$, $S^2\hookrightarrow \CPP^3\rightarrow S^4$, cf.~\cite{Medina:2002pc,Dolan:2003kq,Abe:2004sa}. Second, a purely group theoretic approach was pursued in~\cite{Guralnik:2000pb,Ramgoolam:2001zx}. 

The Hilbert space in both approaches agrees with the Hilbert space we
found from a generalization of Berezin-Toeplitz quantization. The
point at which the approaches differ is in the handling of {\em radial
  fuzziness}. As we showed above, the algebra of quantum operators
$\hat{x}^\mu$ exhausts all of $\sEnd(\CCH_k)$. Therefore the algebra
of quantum operators is isomorphic to the algebra of lower Berezin
symbols of the complex projective space $\CPP^r$ used in the embedding
$S^d\, \embd\, \CPP^r$, and not to the corresponding algebra for
$S^d$. This means that at quantum level the multiplication of two
quantized functions yields modes which should be interpreted as
transverse or radial to the embedding $S^d\, \embd\, \CPP^r$.

There are now two solutions to this problem in the literature. In~\cite{Grosse:1996mz,Guralnik:2000pb,Ramgoolam:2001zx} it was suggested to project out these modes after operator multiplication, which yields a nonassociative algebra. In~\cite{Medina:2002pc}, where fuzzy $S^4$ was used as a regulator for quantum field theories, it was suggested to modify the Laplace operator such that the unwanted modes are dynamically punished by a mass term, i.e.\ their excitation is suppressed.

As eliminating the radial modes by projecting them out after multiplication immediately yields inconsistencies in the interpretation of solutions to the Basu-Harvey equation in terms of fuzzy 3-spheres (see e.g.~\cite{Nastase:2009ny}), we insisted on keeping these modes. This allowed us to interpret fuzzy $S^3$ and fuzzy $S^4$ as quantizations of Nambu-Poisson manifolds under the assumption of a reasonable correspondence principle. 

Note also that the $d$-Lie bracket indeed vanishes if one of the arguments is a purely radial mode. Moreover, the $d$-Lie bracket always yields operators which are quantizations of a function on $S^d$. That is, if one uses exclusively $d$-Lie brackets and avoids the binary operator product, the radial modes are naturally projected out.

\subsection{Fuzzy scalar field theory on $S^d$}

Recall that since our construction respects the isometries of the
sphere, we have the integration formula 
\begin{equation}\label{IntegralFormula}
 \int_{S^d}\, \dd\mu_{S^d}~ f=\frac{1}{\rm vol}~\int_{\CPP^r}\,
 \dd\mu~ \rho(f)=\frac{{\rm vol}(S^d)}{k}\,\tr(\hat{f}\, )~.
\end{equation}
Here ${\rm vol}$ is some constant volume factor, $\rho(f)$ the image of $f\in\CC^\infty(S^d)$ in $\CC^\infty(\CPP^r)$, ${\rm vol}(S^d)$ is the volume of $S^d$, and $\hat{f}$ is the quantization of $f$. This follows since the Rawnsley $\eps$-function is constant for $\CPP^r$ together with $\sSO(d+1)$ invariance of the embedding $S^d\,\embd\, \CPP^r$. One could contemplate replacing $\fh$ by $Q(\sigma_R(\fh\,))$ in \eqref{IntegralFormula}, thereby projecting out radial fuzziness before integrating. At the moment, this seems to us to be merely a matter of taste.

The Laplace operator on $S^d$ is given by the restriction of the
Laplace operator on $\CPP^r$. Recall that the space of global
holomorphic sections $H^0(\CPP^r,\CO(k))$ consists of homogeneous
polynomials of degree $k$. This space carries an irreducible
representation of $\sSU(r+1)$, as well as the spinor representation of
$\sSO(d+1)$. The Laplace operators on $\CPP^r$ and $S^d$ act as the
quadratic Casimir operators of the respective isometry groups in these representations.
At quantum level, the Laplace operator on $S^d$ is obtained by
either the Berezin push or the Berezin-Toeplitz lift of the restricted
Laplace operator in the continuum. Thus although we allow modes which do not correspond to hyperspherical harmonics to arise from the operator product, by this definition we do not allow them to propagate. 

The final question concerns the definition of the functional integration
measure. Contrary to the approach of~\cite{Medina:2002pc}, which was
optimized for numerical purposes, we restrict the space of operators
to the set $\Sigma_R$ in the domain of integration, which is the
intersection of the space of quantizable functions on $\CPP^r$ with
the space of hyperspherical harmonics on $S^d$.
Thus instead of projecting out radial modes, we distinguish the
quantum spheres from quantum projective spaces by a different Laplace
operator, possibly by a different integration formula, and by additionally choosing a different functional integration measure when constructing quantum field theories on these fuzzy spheres.

\section{Quantization of hyperboloids\label{Hyperboloids}}

Our approach to quantizing spheres $S^d$ was based on properties of the
euclidean Clifford algebra $Cl(\FR^{d+1})$. A natural question at this stage is
whether it is possible to extend our quantization procedure using
Clifford algebras for indefinite metrics. The answer is affirmative if
we relax our quantization axiom Q1 and allow for non-unitary representations.

\subsection{Classical hyperboloids}

Recall that a space-like direction is turned into a time-like one by
multiplying the Clifford algebra generator $\gamma^\mu$ corresponding to this
direction by $\di$. In this way we obtain the spinor representation of
the isometry group of the space $\FR^{p,q}$ of dimension $d+1:=p+q$. Into
this space we can embed the $d$-dimensional hyperbolic space
$H^{p,q}$ as the quadric
\begin{equation}\label{embeddingH}
 x^\mu \,x^\nu\, \eta_{\mu\nu}:=(x^1)^2+\dots +(x^p)^2-(x^{p+1})^2-\dots -(x^{p+q})^2=r~,
\end{equation}
where $\eta_{\mu\nu}$ is the metric on $\FR^{p,q}$. We will always
consider the case $r>0$. This restriction eliminates only cones, as by
multiplying the embedding equation by $-1$ one exchanges the r{o}les
of $(p,q)$ and inverts the sign of the curvature. The hyperboloid $H^{p,q}$ corresponds to the coset $\sSO(p,q)/\sSO(p-1,q)$, and $H^{d+1,0}=S^{d}$. For $p=1$, the hyperboloid splits into two sheets. 

The treatment of hyperboloids proceeds analogously to the analysis
of spheres. An embedding into $\FR^{p+q}$ is obtained by substituting
trigonometric functions with hyperbolic functions in
\eqref{sphere.coordinates}, as appropriate for angles in a plane of
signature $(1,1)$, and setting $R=\sqrt r$. The same substitution
applies to the volume element \eqref{sphere.vol}. The natural Nambu
brackets differ from those on the sphere only through the
volume element that one divides by, and we thus define the Nambu bracket on $H^{p,q}$ by
\begin{equation}
 \{f_1,\ldots,f_d\}:=\frac{\eps^{i_1\ldots i_d}}{{\rm
     vol}_{\varphi}}\, \derr{f_1}{\varphi^{i_1}}\ldots\derr{f_d}{\varphi^{i_d}}~,
\end{equation}
which translates into the Nambu bracket of the embedding coordinates
\begin{equation}
 \{x^{\mu_1},\dots ,x^{\mu_d}\}=R^{d-1}\,\eps^{\mu_1\dots \mu_d}{}_{\mu_{d+1}}
 \, x^{\mu_{d+1}}~.
\end{equation}
Here we have defined $\eps^{\mu_1\dots
  \mu_d}{}_{\mu_{d+1}}:=\eps^{\mu_1\dots \mu_d\nu}\, \eta_{\nu\mu_{d+1}}$. 

\subsection{Quantization of $H^{p,q}$}

As we are concerned only with the kinematical problem of quantization,
which we presume to lead to an algebra of quantized functions
approximating the algebra of functions on a space in a well-defined
manner, we can choose to relax the quantization axiom Q1 by mapping
real functions to non-hermitian operators and thus to work with
non-unitary representations. This was done in~\cite{Fakhri:2003cu} in order to
construct a fuzzy $AdS_2$. This approach is a straightforward
generalization of the description of quantum spheres given in
Section~\ref{Qspheres}, and it also fits into the deformation
quantization prescription of Section~\ref{NPquant} For a quantization of a hyperboloid using unitary representations, see e.g.~\cite{Bak:2002rq}.

To allow for an indefinite metric in the Clifford algebra, we have to
allow for non-hermitian generators.\footnote{Recall that the square of a
  hermitian matrix always has positive eigenvalues.} To quantize the
hyperboloid $H^{p,q}$ embedded in $\FR^{p,q}$, we thus multiply the
generators $\gamma^\mu$ along the time-like directions $\mu=p+1,\dots
,p+q$ by a factor of $\di$ and follow the same steps as in the
quantization of the sphere $S^{p+q-1}$. The factors of $\di$ guarantee
that the equation $\xh^\mu\, \xh^\nu\, \eta_{\mu\nu}=R^2_F\, \unit_{\CCH_k}$ is satisfied for the indefinite metric $\eta_{\mu\nu}$. We introduce again the $d$-Lie algebra bracket by the lift of the truncated Nambu-Poisson structure on the set of lower Berezin symbols to the operator algebra. It is only for $d\leq 4$ that this bracket agrees with the totally antisymmetric operator product
\begin{equation}
  {}[\xh^{\mu_1},\dots ,\xh^{\mu_d}]= -\di\,\hbar\,R^{d-1}\,\eps^{\mu_1\dots
    \mu_d}{}_{\mu_{d+1}}\, \xh^{\mu_{d+1}}
\end{equation}
at linear level. This bracket on its own forms the $d$-Lie algebra $A_{p,q}$. Recall that every simple $d$-Lie algebra over $\FR$ is isomorphic to a $d+1$-dimensional $d$-Lie algebra $A_{p,q}$, for some $(p,q)$ with $d=p+q-1$, cf. e.g.~\cite{Figueroa-O'Farrill:arXiv0805.4760}.

As the technical details of the construction (e.g.\ the restriction to
certain irreducible representations for odd-dimensional hyperboloids)
work exactly as for spheres, we refrain from going into further
details. One should stress, however, that while the quantization of
spheres is intimately related to harmonic analysis in the sense that $\sEnd(\CCH_k)$ was related to certain hyperspherical harmonics, this is not the case for the quantum hyperboloids. Thus their quantization is somewhat different in spirit from the standard examples of noncommutative spaces, such as the noncommutative torus. 

Strictly speaking, we actually quantize the one-point
compactifications of the hyperboloids, as there is still an embedding of this compactified hyperboloid into the complex projective space appearing in the construction. Here a point $\varphi=(\varphi^1,\dots,\varphi^d)$ on $H^{p,q}$ is mapped to a point $\varphi'$ on the sphere $S^d$ with the same angular coordinates and subsequently embedded into $\CPP^r$ via the Jordan-Schwinger transform \eqref{JordanSchwingerTrafo}. In this embedding, the point corresponding to infinity on the hyperboloid is also mapped to a point of $S^d$. It is in this sense that we quantize the compactifications of the hyperboloids.

\section{Quantization of superspheres}

One can further extend our approach to the quantization of spheres to superspheres. As before, one constructs an embedding into some complex projective superspace, whose Berezin-Toeplitz quantization induces a quantization on the embedded supersphere.  A natural guess would be to try to use the Clifford-Weyl algebra to perform the embedding.  However, the Weyl subalgebra admits no finite-dimensional representations, and thus would require a projective superspace with infinite fermionic dimensions. We therefore use another approach.

\subsection{Fuzzy projective superspaces}

The Berezin-Toeplitz quantization of complex projective superspace $\CPP^{m|n}$ is discussed in detail in~\cite{Murray:2006pi,IuliuLazaroiu:2008uu}. On $\CPP^{m|n}$, there are homogeneous coordinates ${Z}_A=(z_\alpha,\zeta_a)\sim (\lambda\, z_\alpha,\lambda\,\zeta_a)$, for any $\lambda\in\FC^\times$, where $z_\alpha$, $\alpha=0,1,\dots ,m$ and $\zeta_a$, $a=1,\dots ,n$ are the bosonic and fermionic (Gra{ss}mann) coordinates, respectively. The quantization of $\CPP^{m|n}$ follows along the same lines as that of $\CPP^n$. The space of global holomorphic sections of the quantum line bundle $\CO(k)$ over $\CPP^{m|n}$ is spanned by homogeneous polynomials of degree $k$ in the $Z_A$, which we identify with the $k$-particle Hilbert subspace of the Fock space of $m+1$ bosonic and $n$ fermionic harmonic oscillators. Their creation and annihilation operators satisfy the superalgebra $\lsc\bfah_A,\bfah_B^\dagger\rsc=\delta_{AB}$, where $\lsc-,-\rsc$ denotes the supercommutator and $\bfah_A,\bfah^\dagger_A$ stands for both bosonic and fermionic creation and annihilation operators depending on the value of the combined index $A$. Our Hilbert space $\CCH_k$ is thus spanned by the vectors
\begin{equation}
\hat{\textbf{a}}_{A_1}^\dag\cdots \hat{\textbf{a}}_{A_k}^\dag|0\rangle~.
\end{equation}
The coherent state vectors for this space are supersymmetric generalizations of the truncated Glauber vectors on $\mathbb{C}^{m+1|n}$. Thus one has
\begin{align}
|Z,k\rangle=\frac{1}{k!}\,\big(\bar{Z}_A\,\hat{\textbf{a}}_A^{\dag}\big)^k|0\rangle~,
\end{align}
and the supersymmetric coherent state projector is given by
\begin{align}
P_Z:=\frac{|Z,k\rangle \langle Z,k|}{\langle Z,k|Z,k\rangle}~.
\end{align}

\subsection{Super-Nambu brackets and $n$-Lie superalgebras}

The quantization of superspheres will rely on a super-Nambu-Poisson structure as introduced in e.g.~\cite{Sakakibara:0208040}. We start from a (split) supermanifold $\CM$ and introduce the convention that a tilde over an expression refers to its $\RZ_2$-grading. A super-Nambu-Poisson structure on $\CM$ is an $n$-ary bracket $\CC^\infty(\CM)^{\otimes n}\rightarrow \CC^\infty(\CM)$ satisfying supersymmetric generalizations of the Leibniz rule and the fundamental identity. The super-Nambu bracket itself carries a $\RZ_2$-grading. Let $\varpi$ be the Nambu-Poisson tensor on $\CM$ and $f_i\in \CC^\infty(\CM)$. The parity of a super-Nambu $n$-bracket is then related to the parity of the Nambu-Poisson tensor according to
\begin{align}
\widetilde{\{f_1,\dots ,f_n\}}=\tilde{\varpi}+\sum^{n}_{i=1}\, \tilde{f}_i~.
\end{align}
The {\em super-Leibniz rule} is now given by
\begin{align}
\{g\,h,f_2,\dots ,f_n\}=g\,\{h,f_2,\dots ,f_n\}+(-1)^{(\sum^{n}_{i=2}\,\tilde{f}_i)\, \tilde{h}}\, \{g,f_2,\dots ,f_n\}\, h~,
\end{align}
the {\em super-fundamental identity} reads as
\begin{align}
\{f_1,\dots ,f_{n-1},\{g_1,\dots ,g_n\}\}&=\sum^{n}_{i=1}\, (-1)^{(\tilde{\varpi}+\sum^{n-1}_{j=1}\, \tilde{f}_j)\, (\sum^{i-1}_{k=1}\, \tilde{g}_k)}\, \{g_1,\dots ,\{f_1,\dots ,f_{n-1},g_i\},\dots ,g_n\}
\end{align}
and the $\mathbb{Z}_2$-graded skewsymmetry relations are 
\begin{align}
\{f_1,\dots, f_i,f_{i+1},\dots ,f_n\}=-(-1)^{\tilde{f}_i\,\tilde{f}_{i+1}}\, \{f_1,\dots, f_{i+1},f_i,\dots, f_n\}~,
\end{align}
where $f_i,g,g_i,h\in\CC^\infty(\CM)$. Any supermanifold $\CM$ equipped with such a bracket is called a {\em Nambu-Poisson supermanifold} and is said to have a {\em super-Nambu-Poisson structure}.

Note that the truncation of a super-Nambu-Poisson algebra can be performed completely analogously to Section \ref{sec.truncNambu} and yields an $n$-Lie superalgebra.

To describe superspheres with a super-Nambu-Poisson structure, we again embed them into cartesian superspace first. Thus we consider $S^{d|c}\, \embd\, \FR^{d+1|c}$ with coordinates $X^M=(x^\mu,\xi^m)$, $\mu=1,\dots ,d+1$, $m=1,\dots ,c$, where $x^\mu$ and $\xi^m$ are the bosonic and fermionic coordinates, respectively. The embedding is given by the equation\footnote{From this equation it follows that we must work in the category of supernumbers as $x^\mu$ cannot be purely real.}
\begin{equation}
 x^\mu\, x^\mu+\xi^m\,\xi^m=R^2~.
\end{equation}
The natural super-Nambu $d+c$-bracket in these coordinates is given by
\begin{equation}
\big\{X^{M_1},\dots ,X^{M_{d+c}} \big\}=R^{d+c-1}\,\Xi^{M_1\dots M_{d+c+1}} \, X^{M_{d+c+1}}~,
\label{Asuper}\end{equation}
where the tensor $\Xi$ is totally $\mathbb{Z}_2$-graded skewsymmetric with $\Xi^{\dots M M\dots }=0$ and $\Xi^{1\dots d+c+1}=1$.

In an analogous way, we may equip $n$-Lie algebras with a $\mathbb{Z}_2$-grading.  An \textit{$n$-Lie superalgebra} is a $\mathbb{Z}_2$-graded vector space $\CA$ endowed with a multilinear bracket that is completely $\mathbb{Z}_2$-graded skewsymmetric and satisfies the super-fundamental identity. The $d+c$-Lie superalgebra with bracket (\ref{Asuper}) is the natural supersymmetric extension of the $d$-Lie algebra $A_{d+1}$ considered previously.

\subsection{Embedding superspheres in projective superspace}

Let us first consider superspheres with an even number of fermionic
directions, i.e.\ superspheres of the form $S^{d|2c}$,
$d,c\in\NN$. For simplicity, we work here with complex fermionic
coordinates $\xi^m$, such that $m=1,\dots ,c$ and the embedding
equation is given by
\begin{equation}\label{embsupersphere}
 x^\mu\, x^\mu+\bar{\xi}\,^m\, \xi^m=R^2~.
\end{equation}
With the same conventions as before, we start from the Clifford algebra $Cl(\FR^{d+1})$ with generators
$\gamma^\mu$ and use the embedding into $\CPP^{r|c\,(r+1)}$ given by
\begin{equation}
 x^\mu=\frac{R}{N}\, \bz_\alpha\, \gamma^\mu_{\alpha\beta}\, z_\beta \
 \eand \ \xi^m=\frac{R}{N}\,\bz_\alpha\, \zeta^m_\alpha~,
\end{equation}
where $R$ is the radius of the supersphere and $N$ is a normalization
constant. As a shorthand notation, we define matrices
$\Gamma^M_{AB}=(\gamma^\mu_{AB},g^m_{AB})$ which allow us to write
\begin{equation}\label{main.embedding.superspheres}
 X^M=\frac{R}{N}\, \bar{Z}_A\, \Gamma^M_{AB}\, Z_B~.
\end{equation}

The normalization $N$ is determined by \eqref{embsupersphere} and the expansion
\begin{equation}
 \sqrt{x^\mu\, x^\mu+\bar{\xi}\,^m\, \xi^m}=\frac RN\,\big(\bz_\alpha\,
   z_\alpha +\dots \big) ~,
\end{equation}
where the ellipsis denotes nilpotent terms. This expression is indeed
well-defined. Recall that one can decompose any supernumber $\lambda$
into complex and nilpotent parts, called its body and soul $\lambda_B$
and $\lambda_S$. Because of the nilpotency of $\lambda_S$, the Taylor
series expansion of $\sqrt{\lambda}$ terminates and is given by
\begin{equation}
 \sqrt{\lambda}=\sqrt{\lambda_B+\lambda_S}=\sqrt{\lambda_B}+\frac{\lambda_S}{2\, \sqrt{\lambda_B}}+\dots ~.
\end{equation}
We thus see that the soul part of the embedding is somewhat
arbitrary. We will exploit this arbitrariness and adapt it to
guarantee the validity of the quantization axiom
Q3\pr. For example, in the case of $S^{2|2c}$ the embedding 
\begin{equation}\label{modified.embedding}
 x^\mu=\frac{R}{N}\,\left(\bz_\alpha\, \sigma^\mu_{\alpha\beta}\,
   z_\beta-\bs\,^m_\alpha\, \sigma^\mu_{\alpha\beta}\, \zeta^m_\beta\right)
 \ \eand \ \xi^m=\frac{R}{N}\,\bz_\alpha\, \zeta^m_\alpha
\end{equation}
yields $N= \bz_\alpha\, z_\alpha+\bs\,^m_\alpha\, \zeta^m_\alpha +\dots$.

Superspheres with an odd number of fermionic dimensions are obtained from those above by reducing one fermionic variable to its real part.
Given the embedding, the quantization is now straightforward. We describe the two examples $S^{2|2}$ and $S^{2|3}$ in detail below.

\subsection{Quantization of $S^{2|2}$ and $S^{2|3}$}

Consider the embedding $S^{2|2}\, \embd\, \CPP^{1|2}$ as given in
\eqref{modified.embedding}. The matrices $\Gamma^M_{AB}$ defined via \eqref{main.embedding.superspheres} are given by 
\begin{align}
\Gamma^M=\left(\left(\begin{array}{c|c}\sigma^\mu_{\alpha \beta} & 0
      \\\hline 0 & \sigma^\mu_{\alpha \beta}\end{array}\right) \ , \
  \left(\begin{array}{c|c}0 & 0 \\\hline \unit_2 & 0\end{array}\right)
  \ , \ \left(\begin{array}{c|c}0 & \unit_2 \\\hline 0 & 0\end{array}\right)\right)~.
\end{align}
The space of spherical harmonics of degree $\leq k$ on the supersphere is spanned by functions of the form
\begin{align}
\Gamma^{M_1}_{A_1 B_1}\cdots \Gamma^{M_j}_{A_j B_j}\, \delta_{A_{j+1}
  B_{j+1}}\cdots \delta_{A_k B_k}\, \bar{Z}_{A_1}\cdots
\bar{Z}_{A_k}\, Z_{B_1}\cdots Z_{B_k}~,~~~0\leq j\leq k~.
\end{align}

The quantization of $S^{2|2}$ builds upon the quantization of
$\CPP^{1|2}$ with quantum line bundle $L=\CO(k)$, and we identify as
always the Hilbert space $\CCH_k$ with $H^0(\CPP^{1|2},L)$. The
restricted lower Berezin symbol $\sigma_R(\hat{f}\, )$ of an operator
$\hat{f}\in\sEnd(\CCH_k)$ is defined by the $L^2$-projection of the
lower Berezin symbol $\sigma(\hat{f}\, )\in \Sigma\subset
\CC^\infty(\CPP^{1|2})$ onto $\Sigma_R \subset \CC^\infty(S^{2|2})$,
where $\Sigma_R$ is the space of quantizable functions on $S^{2|2}$.
The lower Berezin symbol $\sigma_R(\hat{f}\, ) \in \Sigma_R$ is defined using a restricted coherent state projector
\begin{align}
P^R_{Z}& :=\sum^k_{m=0}\, X^{M_1}\cdots
X^{M_m}\,k!\,\left(\frac{2}{R}\right)^m\, \Gamma^{M_1}_{A_1B_1}\cdots
\Gamma^{M_m}_{A_m B_m}\nonumber \\ & \qquad \qquad \times\, \hat{\textbf{a}}_{A_1}^\dag\cdots
\hat{\textbf{a}}_{A_m}^\dag\,\hat{\textbf{a}}_{C_1}^\dag\cdots
\hat{\textbf{a}}_{C_{k-m}}^\dag|0\rangle \langle
0|\hat{\textbf{a}}_{B_1}\cdots
\hat{\textbf{a}}_{B_m}\,\hat{\textbf{a}}_{C_1}\cdots
\hat{\textbf{a}}_{C_{k-m}}~.
\end{align}
The quantization prescription then gives
\begin{align}
X^M ~\longmapsto~ \hat{X}^M=\frac R{k!}\, \Gamma^{M}_{AB}\,
\hat{\textbf{a}}_{A}^\dag\, \hat{\textbf{a}}_{C_1}^\dag\cdots
\hat{\textbf{a}}_{C_{k-1}}^\dag|0\rangle\langle
0|\hat{\textbf{a}}_{B}\, \hat{\textbf{a}}_{C_1}\cdots \hat{\textbf{a}}_{C_{k-1}}~.
\end{align}
The bracket of the $4$-Lie superalgebra obtained from the truncated super-Nambu-Poisson structure satisfies the quantization axiom Q3\pr\ again by definition and at linear level agrees with the totally super-skewsymmetric operator product 
\begin{equation}
 \hat{X}^{\lsc M_1}\,\hat{X}^{M_2}\,\hat{X}^{M_3}\,
 \hat{X}^{M_4\rsc}=-\di\, \hbar \, R^3 \, \Xi^{M_1M_2M_3M_4M_5}\, \hat{X}^{M_5}~,
\end{equation}
where $\lsc - \rsc$ denotes the total $\mathbb{Z}_2$-graded
skewsymmetrization of the enclosed indices. (Here one has to translate
$\xi^m$ back to real coordinates.) This follows from
$g^1=g^2=\diag(\di\, \unit_2,-\di\, \unit_2)$,
$[g^1,\gamma^\mu]=[g^2,\gamma^\mu]=0$, and $\gamma^1\, \gamma^2\, \gamma^3\, g^1=\diag(\unit_2,-\unit_2)=-\di\, g^2$.

The supersphere $S^{2|3}$ is obtained from the embedding $S^{2|3}\,
\embd\, \CPP^{1|4}$. It is given by the matrices
\begin{align}
&\Gamma^M=\left(\left(\begin{array}{c|c|c}\sigma^\mu_{\alpha \beta} & 0 & 0 \\\hline  0& \sigma^\mu_{\alpha \beta} & 0 \\\hline 0 & 0 & \sigma^\mu_{\alpha \beta}\end{array}\right)\,,\,\left(\begin{array}{c|c|c}0 & \unit_2 & 0 \\\hline 0 & 0 & 0 \\\hline 0 & 0 & 0\end{array}\right)\,,\,\left(\begin{array}{c|c|c}0 & 0 & 0 \\\hline \unit_2 & 0 & 0 \\\hline 0 & 0 & 0\end{array}\right)\,,\,\left(\begin{array}{c|c|c}0 & 0 & \unit_2 \\\hline 0 & 0 & 0 \\\hline \unit_2 & 0 & 0\end{array}\right)\right)~,
\end{align}
where the blocks correspond to the splitting of the homogeneous coordinates $Z_A$ on $\CPP^{1|4}$ according to $(z_\alpha,\zeta^1_\alpha,\zeta^2_\alpha)$. Note that the super-Nambu bracket here is of odd parity. The remainder of the quantization follows easily from the considerations above.

\section{Quantization of $\FR^n$ by foliations}

As our final set of examples, we will now look at the implications of
our quantization axioms for the quantization of $\FR^n$. The relevant
$n$-Lie algebras at linear level correspond to Nambu-Heisenberg
$n$-Lie algebras, which in turn suggest a quantization of $\FR^n$ in terms of
foliations by fuzzy spheres $S_F^{n-1}$ or
noncommutative hyperplanes $\FR^{n-1}_\theta$. We also briefly study
an extension of this quantization by adding an extra outer
automorphism to the Nambu-Heisenberg $n$-Lie algebra, which describes
a twisting of the $n$-Lie algebra and a dimensional oxidation of the
quantization of $\FR^n$.

\subsection{Nambu-Poisson structures on $\FR^n$ and Nambu-Heisenberg $n$-Lie algebras}

The natural Nambu $n$-bracket on $\FR^n$ is defined by the linear
extension (via the generalized Leibniz rule) and completion (with
respect to the canonical $L^2$-norm) of the bracket
\begin{equation}\label{NPS.can.Rn}
 \{x^{\mu_1},\dots ,x^{\mu_n}\}=\eps^{\mu_1\dots \mu_{n}} \ .
\end{equation}
This Nambu-Poisson structure is naturally
$\sSO(n)$-invariant. Additionally, one can impose further
Nambu-Poisson structures on $\FR^n$ with Nambu $n-1$-brackets. The
$\sSO(n)$ symmetry suggests to add the Nambu-Poisson structure of a foliation of
$\FR^n$ by spheres\footnote{In the case of $\FR^{p,q}$, one would
  instead use the hyperboloids $H^{p,q}$.} $S^{n-1}$, with bracket
\begin{equation}
 \{x^{\mu_1},\dots ,x^{\mu_{n-1}}\}=R^{n-2}\,\eps^{\mu_1\dots\mu_{n-1} \mu_n}\, x^{\mu_n}~.
\end{equation}
Alternatively, one could break the $\sSO(n)$ invariance to $\sSO(n-1)$
and introduce the
Nambu-Poisson structure of a foliation of $\FR^n$ by hyperplanes
$\FR^{n-1}$, with bracket
\begin{equation}
\big\{x^{\check{\mu}_1},\dots ,x^{\check{\mu}_{n-1}} \big\}=\eps^{\check{\mu}_1\dots \check{\mu}_{n-1}}~,~~~\check{\mu}_i=1,\dots ,n-1~.
\end{equation}
In the latter case, we can continue and introduce additionally a
Nambu-Poisson structure with a Nambu $n-2$-bracket, and so on. We
denote the space $\FR^n$ endowed with $k\leq n-2$ successive hyperplane
foliations and one spherical foliation by $\FR^n_k$. In the case $k=n-2$ there is no spherical
foliation, while for $k=0$ there is only the spherical foliation.

The components of the Nambu-Poisson tensor are constants, so that the truncation of the Nambu-Poisson structure as presented in Section \ref{sec.truncNambu} unfortunately does not work here. We will therefore restrict to an $n$-Lie algebra structure which is nontrivial only at linear level and there agrees with the totally antisymmetric operator product. Correspondingly, the quantization axiom Q3\pr\ can only be satisfied at linear level. Thus, the Nambu-Poisson structure
\eqref{NPS.can.Rn} has to turn under quantization into the $n$-Lie
algebra $\CA_{\rm NH}$ with bracket 
\begin{equation}
{}[\xh^{\mu_1},\dots
,\xh^{\mu_n}]=-\di\,\hbar\,\eps^{\mu_1\dots \mu_{n}}\, \unit~,
\end{equation}
where the vector space $\CA_{\rm NH}$ is spanned by the operators
$\xh^\mu$, $\mu=1,\dots,n$, and $\unit$. This algebra is called
the {\em Nambu-Heisenberg $n$-Lie algebra}. The nested foliations
yield additional $n-1$-Lie algebra structures on $\CA_{\rm NH}$. We will study these quantizations in the following, starting from the quantizations of $\FR^3_0$ and $\FR^3_1$.

\subsection{Quantization of $\FR_0^3$ and $\FR_1^3$}

The 3-Lie algebra $\CA_{\rm NH}$ was examined in the original
paper~\cite{Nambu:1973qe}, as well as in~\cite{Takhtajan:1993vr}. It is generated by four elements $\hat{x},\hat{y},\hat{z},\unit$ with the defining relation
\begin{equation}\label{defNH3}
{}[\hat{x},\hat{y},\hat{z}]=-\di\,\hbar\, \unit~.
\end{equation}
This relation is a consistency constraint for a quantization of both $\FR^3_0$ and $\FR^3_1$ according to our generalized quantization axioms.

To realize the quantization map on the endomorphism algebra of some Hilbert space $\CCH$, we assume that the generator $\unit$ appearing on the right-hand side of \eqref{defNH3} is central in this algebra and acts on vectors of the Hilbert space $\CCH$ as multiplication by a complex number. This implies that its commutator with any other endomorphism vanishes. From the definition of the 3-bracket as a totally antisymmetrized operator product,
\begin{equation}\label{3bracketwithouttrace}
 [\hat A,\hat B,\hat C]:=
\left\{\begin{array}{ll}
\hat A\,[\hat B,\hat C]+\hat B\,[\hat C,\hat A]+\hat C\, [\hat A,\hat C]&\mbox{for}~~~\hat A,\hat B,\hat C\in{\rm span}(\xh,\hat{y},\hat{z},\unit)\\
0&\mbox{else}
\end{array}\right.~~~,
\end{equation}
it is clear that a central element of the 2-Lie bracket will not, in general, be a central element in the 3-Lie algebra. Thus we will have the relations
\begin{equation}\label{unit3bracket}
 [\unit,\hat A,\hat B]=\alpha\, [\hat A,\hat B]~,~~~\alpha\in\FC^\times
\end{equation}
for all $\hat A,\hat B$, rather than $[\unit,\hat A,\hat B]=0$. However, if
a 3-Lie algebra satisfying \eqref{defNH3} as well as
\eqref{unit3bracket} is given as an operator algebra with a
finite-dimensional\footnote{This condition is necessary in order to
  avoid issues related to trace-class operators.} representation, we
can construct a new bracket\footnote{It should be stressed that in
  using this bracket, we lose the interpretation of our quantization
  in terms of factoring out ideals in the corresponding universal enveloping algebra.}
\begin{equation*}\label{tracetrick}
 [\hat A,\hat B,\hat C]_{\rm NH} :=
\left\{\begin{array}{ll}
\tr\big(\hat A\, [\hat B,\hat C]+\hat B\, [\hat C,\hat A]+\hat C\, [\hat A,\hat B]\big)~\unit&\mbox{for}~~~\hat A,\hat B,\hat C\in{\rm span}(\xh,\hat{y},\hat{z},\unit)\\
0&\mbox{else}
\end{array}\right.~~~,
\end{equation*}
which trivially generates a 3-Lie algebra structure for which $\unit$
is a 3-central element. We therefore restrict our considerations to
the bracket~\eqref{3bracketwithouttrace}. In fact, below we will necessarily
have to deal with infinite-dimensional Hilbert spaces $\CCH$.

The possibilities of realizing the relation~\eqref{defNH3} as a
totally antisymmetric operator product have been listed in
\cite{Nambu:1973qe}. Nambu employs the Lie algebras of $\sSU(2)$,
$\sSO(1,2)\cong \sSL(2,\FR)$, the euclidean group in two dimensions,
and the galilean group in one dimension. Here we restrict ourselves to
the three-dimensional cases. We will show below that the first three
cases correspond to quantizations of $\FR^3_0$, $\FR^{1,2}_0$, and
$\FR^3_1$, respectively. The generic constraints on three-dimensional
Lie algebras which realize (\ref{defNH3}) and (\ref{unit3bracket}) are
derived in Appendix~\ref{NHLiealg}, where we also derive the most general
form of the Lie algebra $\frg_{\CA_{\rm NH}}$ associated to the
Nambu-Heisenberg 3-Lie algebra $\CA_{\rm NH}$.

\subsubsection*{$\FR_0^3$}

In the first case of $\sSU(2)$, the Lie algebra yielding~\eqref{defNH3} corresponds
to the coordinate algebra of the fuzzy sphere $S_F^2$. The radial
restriction $\xh^\mu\,\xh^\mu=\rho\,\unit_{\CCH}$ for a
constant $\rho\in\FC^\times$, however,
is missing. We thus obtain a foliation of $\FR^3$ by fuzzy
spheres. This space is usually denoted $\FR^3_\lambda$ in the
literature~\cite{Hammou:2001cc,Batista:2002rq}. Recall that on a fuzzy sphere built on the Hilbert space $\CCH_k=H^0(\CPP^1,\CO(k))$, the 3-bracket is given by
\begin{equation}
{}[\xh^1,\xh^2,\xh^3]=\left(\frac{R}{k!}\right)^3\,
\big((k-1)!\big)^2\,k\, 
(\eps^{\mu\nu\kappa}\,\sigma^\mu\,\sigma^\nu\,
\sigma^\kappa)_{\alpha\beta}\, 
\st{\alpha}{\beta}=-\di\,\frac{6R^3}{k}\,\unit_{\CCH_k}~,
\label{fuzzy3bracket}\end{equation}
and the fuzzy radius is $R_F=R_{F,k}:=R\, \sqrt{1+\frac{2}{k}}$. As
$R_{F,k}^2\,\unit_{\CCH_k}=\xh^\mu\, \xh^\mu$ is not fixed, the
relation \eqref{defNH3} admits fuzzy spheres of various radii. For
given deformation parameter $\hbar$, we have $\hbar=\frac{6R^3}{k}$
from (\ref{fuzzy3bracket}) and consequently a quantization of the radius of the fuzzy sphere
\begin{equation}
R_{F,k}=\sqrt{1+\frac{2}{k}}~ \sqrt[3]{\frac{\hbar\, k}{6}}
\end{equation}
built on the Hilbert space $\CCH_k$. 

Let us introduce now the Hilbert space $\CCH:=\bigoplus_{k\in\NN}\,
\CCH_k$ together with the algebra of ``quantum functions''
$\CA:=\bigoplus_{k\in\NN}\, \sEnd(\CCH_k)$. This corresponds to a
``discrete foliation'' of $\FR^3$ by fuzzy spheres with radii
$R_{F,k}$. The quantization of a polynomial in the coordinates $x^\mu$
corresponding to a function on $\FR^3$ is given by a quantization of
this coordinate function on each fuzzy sphere. The 3-bracket is
non-vanishing only on those elements of $\CA$ which are all at most linear elements
of the same subalgebra $\sEnd(\CCH_k)$. The geometry corresponding to
the noncommutative algebra of functions $\CA$ is the space
$\FR^3_\lambda$, with $\lambda=\sqrt{2\hbar/3R}$. An explicit star
product (\ref{starprod}) is constructed in~\cite{Hammou:2001cc} using
the embedding $\FR^3\hookrightarrow\FC^2$ and the coherent state
(Wick-Voros) star product on noncommutative $\FC^2$.

Let us now examine how the associated Lie algebra $\frg_\CA$ is related to the isometries of $\FR^3_\lambda$. A priori, there is no reason to expect a direct connection, as the ``fundamental'' object in this quantization is the Lie bracket of the quantized coordinate functions $\xh^\mu$. The associated Lie algebra of this 2-Lie algebra is the 2-Lie algebra itself, i.e.\ $\asu(2)$, which indeed corresponds to the (continuous) isometries of $\FR^3_\lambda$. 

The associated Lie algebra $\frg_\CA$ is of dimension six with generators $D_{\mu\nu}:=D(\xh^\mu\wedge\xh^\nu)$, $\mu,\nu=0,1,2,3$, where $\hat x^0:=-\di\,\hbar\,\unit$. In the basis
\begin{equation}
\begin{aligned}
X^1=D_{12}-D_{30}~,~~~X^2&=D_{23}-D_{10}~,~~~X^3=D_{13}+D_{20}~,\\[4pt]
Y^1=D_{12}+D_{30}~,~~~Y^2&=D_{23}+D_{10}~,~~~Y^3=D_{13}-D_{20}~,
\end{aligned}
\end{equation}
the non-vanishing commutation relations read (see Appendix~\ref{NHLiealg})
\begin{equation}
{}[X^i,X^j]=2\eps^{ijk}\,X^k~,~~~[Y^i,Y^j]=2\eps^{ijk}\, Y^k~,~~~i,j,k=1,2,3~.
\label{R03Liealg}\end{equation}
Thus the associated Lie algebra is $\mathfrak{so}(3) \oplus \mathfrak{so}(3)$, as expected since $\CA\cong A_4$ in this case. The generators $X^i-Y^i$ generate the $\asu(2)$ isometries on $\FR^3_\lambda$. The remaining generators transform the operator $\rho\,\unit$, which corresponds to a (scalar) radius function in the geometric picture. Although they describe non-geometric symmetries, their appearance is very natural if we use the PBW isomorphism (\ref{PBWiso}) to identify $\CA$ with the universal enveloping algebra $U_2(\asu(2))$. As discussed in~\cite{Batista:2002rq}, since $U_2(\asu(2))$ is a Hopf algebra it has a natural \emph{quantum} isometry group given by the Drinfel'd quantum double $D(U_2(\asu(2)))$, which in this case is the crossed product of $U_2(\asu(2))$ with the coordinate algebra $\FC(\sSU(2))$ of the $\sSU(2)$ Lie group by the coadjoint action of $U_2(\asu(2))$ on $\FC(\sSU(2))$. Here $U_2(\asu(2))$ acts on itself by the left adjoint action and corresponds to the geometric symmetries above, while $\FC(\sSU(2))$ acts on $U_2(\asu(2))$ by the right coregular action and corresponds to the non-geometric symmetries.

\subsubsection*{$\FR_0^{1,2}$}

An analogous construction holds for the 3-bracket built on the Lie
algebra $\sSO(1,2)\cong \sSL(2,\FR)$. Here the fuzzy spheres are
replaced by the fuzzy hyperboloids $H_F^{1,2}$ (or $H_F^{2,1}$)
constructed in Section~\ref{Hyperboloids} This
defines the noncommutative space $\FR^{1,2}_\lambda$. We thus obtain a foliation of $\FR^3$ by fuzzy hyperboloids in this case.

\subsubsection*{$\FR_1^{3}$}

In the third case, the euclidean group in two dimensions, we start from the Lie algebra
\begin{equation}\label{defE2}
[\xh^1,\xh^2]=-\di\,\xi\, \xh^3~,~~~[\xh^3,\xh^1]=[\xh^3,\xh^2]=0
\end{equation}
with a constant $\xi\in\FC$. This algebra breaks the explicit $\sSO(3)$ invariance down to $\sSO(2)$. Since $\xh^3$ is a central element of this algebra we can assume it acts as $\alpha\,\unit$, $\alpha\in\FC$ on any irreducible representation, and thus we can put $\xi=\frac{\hbar}{\alpha^2}$. The 3-bracket defined from the antisymmetric operator product is then given by
\begin{equation}\label{E2cond}
 [\xh^1,\xh^2,\xh^3]=\xh^3\,[\xh^1,\xh^2]=-\di\,\hbar\, \unit~.
\end{equation}
The quantum geometry behind this algebra $\CA$ is thus a foliation of $\FR^3$ in terms of standard noncommutative planes\footnote{For a construction of this space via Berezin-Toeplitz quantization, see~\cite{IuliuLazaroiu:2008pk}.} $\FR^2_\theta$ extending in the directions parameterized by $x^1$ and $x^2$. The eigenvalues of $\xh^3$ corresponding to the $x^3$ position of the noncommutative plane determine the noncommutativity parameter $\theta=\frac{\hbar}{x^3}$. This implies that the plane through $x^3=0$ is somewhat ill-defined. As $\sSO(3)$-invariance is broken by the Nambu-Poisson structure here, one can equally well interpret the eigenvalues of $(\xh^3)^{-1}$ as the position of the noncommutative plane. In this case, one obtains a commutative plane $\FR^2$ through the origin. The noncommutative space with coordinate algebra $\CA$ in this case is denoted $\FR_{1,\theta}^3$.

The associated Lie algebra $\frg_\CA$ is again spanned by the six
generators $D_{\mu\nu}:=D(\xh^\mu\wedge\xh^\nu)$, $\mu,\nu=0,1,2,3$
satisfying the non-vanishing commutation relations (see Appendix~\ref{NHLiealg})
\begin{equation}
\begin{aligned}
{}[D_{12},D_{13}]=-D_{10}~,~~~&[D_{10},D_{20}]=-D_{30}~,\\[4pt]
[D_{12},D_{23}]=-D_{20}~,~~~&[D_{10},D_{12}]=-D_{13}~,\\[4pt]
[D_{23},D_{13}]=+ D_{30}~,~~~&[D_{20},D_{13}]=-D_{23}~.
\end{aligned}
\label{R13Liealg}\end{equation}
This is an indecomposable simple Lie algebra. The isometries of $\FR^3_{1,\theta}$, however, span the Lie algebra $\mathbb{R} \oplus \aso(2)$, corresponding to translations along the $x^3$ direction and rotations in the foliating planes. As the $\aso(2)$ rotations act as outer derivations of the Heisenberg algebra $[\xh^1,\xh^2]=-\di\,\theta\,\unit$, there is no relation between the isometries and the associated Lie algebra. Worthy of note is the maximal subalgebra of the associated Lie algebra given by
\begin{equation}
{}[D_{12},D_{23}]=-D_{20}~,~~~[D_{12},D_{20}]=D_{23}~,~~~[D_{30},-]=0~,
\end{equation} 
which is isomorphic to $\mathfrak{iso}(2)\ltimes \mathbb{R}$. We conclude that the associated Lie algebra only describes non-geometric symmetries, and hence purely quantum isometries of the space $\FR^3_{1,\theta}$ in the sense explained above. Note that as the operators appearing in the construction of $\FR^2_\theta$ are not trace-class, we cannot use the trick \eqref{tracetrick} to render $\unit$ a central element of the $3$-Lie algebra of coordinate functions in this case. 

\subsection{Quantum geometry of M5-branes}

We have thus found a geometric interpretation of the equations 
\begin{equation}\label{relationsCS}
 [\hat X^\mu,\hat X^\nu,\hat X^\kappa]=-\di\,\hbar\, \Theta^{\mu\nu\kappa}\,\unit \ \eand \ [\unit,-,-]=0
\end{equation}
found by Chu and Smith in~\cite{Chu:2009iv} describing the quantum geometry of an M5-brane in a constant $C$-field background, where
\begin{equation}
\Theta^{\mu\nu\kappa}=\left\{\begin{array}{ll}\eps^{\mu\nu\kappa}\, C_1~,~~~&\mu,\nu,\kappa=0,1,2\\\eps^{\mu\nu\kappa}\, C_2~,~~~&\mu,\nu,\kappa=3,4,5\\0&\mbox{otherwise}\end{array}\right.
\end{equation}
and $C_1,C_2$ are constants related to the components of the
$C$-field. They correspond to the quantizations of $\FR^{1,2}\times
\FR^3$ with foliations by either fuzzy hyperboloids and spheres or
noncommutative planes. We may heuristically regard the foliating
noncommutative geometries as the dimensional reductions of the
M5-brane configuration in the presence of a $C$-field to a
configuration of D-branes in the appropriate $B$-field background.

\subsection{Quantization of $\FR_k^n$}

Let us now generalize our construction to $n$-Lie algebras. The
Nambu-Heisenberg $n$-Lie algebra is given by
\begin{equation}\label{NHAn}
 [\xh^1,\dots ,\xh^n]=-\di\, \hbar\, \unit~.
\end{equation}
Assuming again that this constraint arises from a quantization of a Nambu $n$-bracket with $n$ coordinates, $\{x^1,\dots ,x^n\}=1$, we are now looking for a quantization of $\FR^n_k$.
To quantize $\FR^n_k$, we have to specify first the number $k$ of nested foliations by noncommutative hyperplanes. The quantization of $\FR^n_0$ corresponds to a quantization via a foliation by $n-1$-dimensional noncommutative spheres, while the quantization of $\FR^n_k$ corresponds to a foliation by noncommutative hyperplanes $\FR^{n-1}_{k-1}$. 
Here we encounter an analogous problem to the correspondence principle not holding even at linear level for $S^d$ with $d>4$. The algebra \eqref{NHAn} only holds on $\FR_k^n$ with $n\leq 4$, as one can verify by direct calculation.

In these constructions, the central operator $\unit$ will not be central in the $n$-Lie algebra. The trick using the trace mentioned above, however, allows us to define an $n$-Lie algebra structure on $\FR^n_0$ where both \eqref{NHAn} and $[\unit,-,\dots ,-]_{\rm NH}=0$ hold. Define a new bracket by the trace over the antisymmetric operator product times the identity as
\begin{equation}
 [\Ah_1,\dots ,\Ah_n]_{\rm NH}:=\left\{\begin{array}{ll}
\tr\big(\eps^{i_1\dots i_n}\,\Ah_{i_1}\cdots \Ah_{i_n}\big)\, \unit&\mbox{for}~~~\Ah_i\in{\rm span}(\xh^1,\ldots,\xh^n,\unit)\\
0&\mbox{else}
\end{array}\right.
\end{equation}
for $n$ odd, and the analogous product with an insertion of $\gamma_\ch$ for $n$ even. This definition yields an $n$-Lie algebra, preserves \eqref{NHAn}, and turns $\unit$ into a central element of the $n$-Lie algebra.

While a dimensional reduction is achieved by reducing an $n$-bracket
to an $n-1$-bracket by filling one of the slots with a selected
generator, the inverse operation of dimensional oxidation can in a
certain sense be realized through an $n$-Lie algebra generalization of Heisenberg
algebras which are twisted by an additional outer automorphism that rotates the
noncommuting coordinates. As this construction involves central elements of $n$-Lie algebras, which are not compatible with the process of factoring ideals out of a tensor algebra to produce a universal enveloping algebra of coordinates, we will have to give up the interpretation in terms of deformation quantization of $n$-Lie algebras in the following.
The {\em twisted Nambu-Heisenberg $n$-Lie algebra} is obtained by introducing an additional generator $\hat J$ and imposing the relations
\begin{equation}\label{NappiWittenn}
 {}[\xh^1,\dots ,\xh^{n}]=-\di\, \hbar\, \unit~,~~~[\hat J,\xh^{\mu_1},\dots ,\xh^{\mu_{n-1}}]=-\di\,\hbar\,\eps^{\mu_1\dots \mu_{n-1}\mu_n}\, \xh^{\mu_n}~,~~~[\unit,-,\dots ,-]=0~.
\end{equation}
One readily verifies that the fundamental identity is indeed
satisfied. In contrast to the Nambu-Heisenberg $n$-Lie algebra, this
$n$-Lie algebra $\CA_{\rm twNH}$ is \emph{metric}, i.e.\ it admits a non-degenerate
$\frg_{\CA_{\rm twNH}}$-invariant inner product in the sense of e.g.~\cite{Figueroa-O'Farrill:arXiv0805.4760}. In fact, for $n>3$ it is the unique indecomposable semisimple lorentzian $n$-Lie algebra of dimension $n+2$~\cite{Figueroa-O'Farrill:arXiv0805.4760}. For $n=3$ it is the semisimple finite-dimensional indecomposable lorentzian 3-Lie algebra obtained by double extension from the compact semisimple Lie algebra $\mathfrak{so}(3)$~\cite{DeMedeiros:2008zm}. For $n=2$ it is the semisimple lorentzian Nappi-Witten Lie algebra~\cite{Nappi:1993ie}, i.e.\ a central extension of the euclidean Lie algebra $\mathfrak{iso}(2)$ in two dimensions.

If the original Nambu-Heisenberg $n$-Lie algebra makes use of the
operator algebra of $S^{n-1}$, the corresponding twisted extension can
be constructed using the operator algebra on $S^{n}$. For clarity, let
us focus on the example $n=3$ and the construction using fuzzy
$S^2$. We embed the Clifford algebra $Cl(\FR^3)$ into the Clifford
algebra $Cl(\FR^4)$ used in the construction of noncommutative
$S^3$. We then put $\hat J=\gamma^4$ and define the 3-bracket
\begin{align}
 [\hat A,\hat B,\hat C]_{\rm twNH}:=&
 \tr\big((\hat A\,[\hat B,\hat C]+\hat B\,[\hat C,\hat A]+\hat
 C\,[\hat A,\hat B])\,\hat J\,\gamma^5\big)\, \unit
 \nonumber \\ 
 &+\, \tr\big((\hat A\, [\hat B,\hat C]+\hat B\, [\hat C,\hat A]+\hat
 C\, [\hat A,\hat B])\, \hat J\,
 \gamma^5\,\gamma^{\mu}\big)\, \gamma^{\mu}
\end{align}
for $\hat A,\hat B,\hat C\in{\rm span}(\xh^1,\xh^2,\xh^3,\hat{J})$, where $\mu,\nu,\kappa=1,2,3$, and $0$ otherwise. This bracket indeed satisfies the relations \eqref{NappiWittenn} for $n=3$. Removing the trace and the projection onto certain Clifford algebra elements, we arrive at the quantum 3-Lie bracket on $S^3$. It is in this sense that we have performed a dimensional oxidation.

The twisted Nambu-Heisenberg $n$-Lie algebra $\CA_{\rm twNH}$ has an associated Lie algebra with $2n$ generators corresponding to two subalgebras. Generators $D_{\mu 0}$ correspond to translations $\FR^n$, and generators $D_{\mu \hat J}$ form an $\mathfrak{so}(n)$ subalgebra. For $n=3$ we find the relations
\begin{align}
[D_{\mu 0},D_{\nu 0}]=0~,~~~[D_{\mu \hat J},D_{\nu \hat J}]=2\eps^{\mu\nu\kappa}\, D_{\kappa0}~,~~~[D_{\mu 0},D_{\nu \hat J}]=2\eps^{\mu\nu\kappa}\, D_{\kappa0}~.
\end{align}
For generic values of $n$, $\frg_{\CA_{\rm twNH}}$ is isomorphic to
the euclidean Lie algebra $\mathfrak{iso}(n)$ in $n$ dimensions. The
quantum isometry groups for $n=2$ and $n=4$ are described
in~\cite{Halliday:2006qc,Meljanac:2008ud}, where explicit star
products (\ref{starprod}) can also be found.

\acknowledgements

We are grateful to C.-S.~Chu, J.~Figueroa-O'Farrill, A.~Konechny, B.~Schroers,
D.~Smith and S.~Vaidya for enlightening discussions. This work was supported by grant ST/G000514/1 ``String Theory
Scotland'' from the UK Science and Technology Facilities Council. The
work of CS was supported by a Career Acceleration Fellowship from the
UK Engineering and Physical Sciences Research Council.

\appendices

\subsection{Extending Nambu-Poisson algebras via the Leibniz rule\label{ExtNPalgebra}}

Given a Nambu-Poisson bracket on a subset $\Upsilon$ of the algebra of
smooth functions $\CC^\infty(\CM)$ on a manifold $\CM$, one can
consistently extend this bracket to the subset
$\FC[\Upsilon]\subset \CC^\infty(\CM)$ of polynomials in elements of
$\Upsilon$. One can use complete induction to verify the fundamental
identity. By direct computation, one readily concludes that the relation
\begin{equation}\label{induction-start}
 \{f_1,\ldots,f_{n-1},\{g_1,\ldots,g_n\}\}=\sum_{i=1}^n\, \{g_1,\ldots,\{f_1,\ldots,f_{n-1},g_i\},\ldots,g_n\}
\end{equation}
implies
\begin{equation}
 \{f_1,\ldots,f_{n-1},\{x \, g_1,\ldots,g_n\}\}=\sum_{i=1}^n\, \{x\, g_1,\ldots,\{f_1,\ldots,f_{n-1},g_i\},\ldots,g_n\}
\end{equation}
for an arbitrary element $x\in\Upsilon$. Furthermore, the relation \eqref{induction-start} implies
\begin{equation}
 \{x \,f_1,\ldots,f_{n-1},\{g_1,\ldots,g_n\}\}=\sum_{i=1}^n\, \{g_1,\ldots,\{x\, f_1,\ldots,f_{n-1},g_i\},\ldots,g_n\}
\end{equation}
as well if and only if
\begin{eqnarray}
&& \sum_{i=1}^n \, \big( \{g_1,\ldots,g_{i-1},x,\ldots,g_n\}\,
\{f_1,\ldots,f_{n-1},g_i\} \nonumber\\ && \qquad\qquad +\,
\{g_1,\ldots,g_{i-1},f_1,\ldots,g_n\}\, \{x,f_2,\ldots,f_{n-1},g_i\}\big)=0~.
\label{iffrel}\end{eqnarray}
The relation (\ref{iffrel}) is satisfied for $f_i,g_i\in\Upsilon$, as here the fundamental identity holds. Moreover, it extends trivially to $\FC[\Upsilon]$ by complete induction. Thus the fundamental identity indeed holds on all of $\FC[\Upsilon]$.

\subsection{Generators of Clifford algebras\label{Cliffalg}}

If $\gamma^{i}$, $i=1,\dots ,2d-1$ generate the Clifford algebra $Cl(\FR^{2d-1})$, then the $2d$-tuple
\begin{equation}
 (\gamma^{\mu})=(\gamma^i\otimes\sigma^2,\unit_s\otimes\sigma^1)~,~~~ s=2^{d-1}~,~~~\mu=1,\dots ,2d
\end{equation}
generates $Cl(\FR^{2d})$. On the other hand, we just add
$\gamma_\ch:=\di^d\, \gamma^1\cdots \gamma^{2d}$ to the generators of
$Cl(\FR^{2d})$ to obtain a set of generators of $Cl(\FR^{2d+1})$. We
can start the induction from the usual Pauli matrices $\sigma^i$,
which generate $Cl(\FR^3)$ and satisfy $[\sigma^i,\sigma^j]=-2\, \di\,
\eps^{ijk}\, \sigma^k$. In this case, all the generators are hermitian
and we have $\gamma_\ch=\diag(\unit_{s},-\unit_{s})$. In the main text,
we use the basis of Pauli matrices given by
\begin{equation}
\sigma^1=\begin{pmatrix} 0&1 \\ 1&0 \end{pmatrix} \ , \qquad
\sigma^2=\begin{pmatrix} 0&\di \\ -\di&0 \end{pmatrix} \ , \qquad
\sigma^3=\begin{pmatrix} 1&0 \\ 0&-1 \end{pmatrix} \ .
\end{equation}

Recall that for even $d+1$, there is a set of generators
$\lambda^a$, $a=1,\dots ,r^2$ of $\au(r)$, $r=2^{\frac{d-1}{2}}$ given by
\begin{equation}
 \frac{1}{\sqrt{r}}\, \unit_r~,~~~\frac{2}{r}\,
 \gamma^\mu~,~~~\frac{2\, \di}{r}\, \gamma^{\mu\nu}~,~~~\frac{2\,
   \di}{r}\, \gamma^{\mu\nu\rho}~,~~~\frac{2}{r}\, \gamma^{\mu\nu\rho\sigma}~,~~~\dots ~,
\end{equation}
where $\gamma^{\mu_1\dots\mu_k}$ is the normalized antisymmetric
  product of gamma-matrices $\gamma^{\mu_1},\dots,\gamma^{\mu_k}$.
With this normalization, they satisfy the Fierz identity
\begin{equation}\label{gammadecompose}
 \lambda^a_{\alpha\beta}\,
 \lambda^a_{\gamma\delta}=\delta_{\alpha\delta}\, \delta_{\beta\gamma}~.
\end{equation}
As these generators of $\au(r)$ form an orthogonal set with respect to
the Hilbert-Schmidt norm, we conclude that all of them are traceless
except for the identity matrix.

\subsection{Tensor product formulas}\label{appTPformulas}

In the main text, we derived that the generators $\gamma^\mu$ of $Cl(\FR^{d+1})$ obey
\begin{equation}
 \gamma^\mu\odot\gamma^\mu=\unit_{r+1}\odot\unit_{r+1}
\end{equation}
in an irreducible representation of $\sSO(d+1)$. Using this result, one readily obtains
\begin{equation}
 -\sum_{\mu,\nu=1}^{d}\,
 \gamma^{\mu\nu}\odot\gamma^{\mu\nu}=(d-2)\,\unit_{r+1}\odot\unit_{r+1}+2\, \gamma_\ch\odot\gamma_\ch~.
\end{equation}
We also find
\begin{equation}
 \begin{aligned}
  \sum_{\mu,\nu=1}^{d}\,
  \gamma_\ch\,\gamma^{\mu\nu}\odot\gamma^{\mu\nu}&=-d\, \gamma_\ch\odot\unit_{r+1}~,\\[4pt]
  (\,\underbrace{\gamma_\ch\odot\unit_{r+1}\odot\ldots\odot\unit_{r+1}}_{\ell}\,
  )^2&=\frac{1}{\ell}\, \big(\unit_{r+1}\odot\cdots\odot\unit_{r+1}+(\ell-1)\, \gamma_\ch\odot\gamma_\ch\odot\unit_{r+1}\odot\cdots\odot\unit_{r+1}\big)~.
 \end{aligned}
\end{equation}

\subsection{Action of
  $\gamma_\ch\odot\gamma_\ch\odot\unit_{r+1}\odot\dots \odot\unit_{r+1}$ on
  $\CV_{k,s}$\label{gammach}}

Consider the quantization of $\CPP^r$, $r=2n-1$ with creation and
annihilation operators satisfying the Heisenberg-Weyl algebra
$[\ah_\alpha,\ah^\dagger_\beta]=\delta_{\alpha\beta}$,
$\alpha,\beta=1,\dots ,2n$. The vectors $\ah^\dagger_\alpha|0\rangle$
generate the reducible spinor representation $V$ of $\sSO(d+1)$, for
$d$ odd. The $k$-fold totally symmetrized tensor product
representation $V^{\odot k}$ is then generated by
$\ah^\dagger_{\alpha_1}\cdots \ah^\dagger_{\alpha_k}|0\rangle$. The
spinor representation $V$ splits into the direct sum of two
irreducible representations, $V=V_+\oplus V_-$, where $V_\pm$ are the
$\pm\,1$ eigenspaces of the chirality operator $\gamma_\ch$. The
totally symmetrized tensor product representations then split according to
\begin{equation}
 \CV_k:=V^{\odot k}=\bigoplus_{s=0}^k\, \big(V_+^{\odot s}\oplus
 V_-^{\odot (k-s)} \big)=:\bigoplus_{s=0}^k\, \CV_{k,s}~.
\end{equation}

We now calculate the action of the operator
$\CO:=\gamma_\ch\odot\gamma_\ch\odot\unit_{r+1}\odot\dots
\odot\unit_{r+1}$ on the subspace $\CV_{k,s}$. For this, split the
creation and annihilation operators into two groups
$(\hat{b}_i,\hat{b}_i^\dagger)=(\ah_i,\ah_i^\dagger)$ and
$(\hat{c}_i,\hat{c}_i^\dagger)=(\ah_{i+n},\ah_{i+n}^\dagger)$, where
$i=1,\dots ,n$. Vectors $|\vec{p},s\rangle \in \CV_{k,s}$ then take
the form $\hat{b}^\dagger_{i_1}\cdots \hat{b}^\dagger_{i_s}\, \hat{c}^\dagger_{i_{s+1}}\cdots \hat{c}^\dagger_{i_k}|0\rangle$ and the operator $\CO$ acts according to
\begin{equation}
\CO|\vec{p},s\rangle=\big(\hat{b}^\dagger_{i_1}\,
\hat{b}^\dagger_{i_2}\unitHH\hat{b}_{i_1}\,
\hat{b}_{i_2}+\hat{c}^\dagger_{i_1}\,
\hat{c}^\dagger_{i_2}\unitHH\hat{c}_{i_1}\, \hat{c}_{i_2}\big)|\vec{p},s\rangle~.
\end{equation}
For a vector $|\vec{p},s\rangle\in\CV_{k,s}$ with $k\geq 3$, we readily verify that $\CO|\vec{p},s\rangle\propto |\vec{p},s\rangle$, and that the eigenvalue of $\CO$ is identical in the representations $\CV_{k,s}$ and $\CV_{k,k-s}$.

\subsection{Nambu-Heisenberg $3$-Lie algebras\label{NHLiealg}}

From (\ref{unit3bracket}) we know that the 3-bracket involving $\hat x^0:=-\di\,\hbar\,\mathbbm{1}$ takes the form
\begin{align}
[\hat x^0,\hat x^\mu,\hat x^\nu]=\alpha\,[\hat x^\mu,\hat x^\nu] \ ,
\label{2Liehatx}\end{align}
where $\alpha \in \mathbb{C}^\times$ and $\mu,\nu=0,1,2,3$. If we rewrite this bracket as
\begin{align}
[\hat x^0,\hat x^\mu,\hat x^\nu]=f^{0\mu\nu}{}^\beta \, \hat x^\beta~,
\end{align}
then the general form of a 3-bracket including $\hat x^0$ can be derived by solving the fundamental identity
\begin{align}
f^{0 \nu \lambda}{}^{\rho}\, f^{\sigma\alpha\rho}{}^{\beta}=f^{\sigma
  \alpha 0}{}^{\rho}\, f^{\rho\nu\lambda}{}^{\beta}+f^{\sigma \alpha
  \nu}{}^{\rho}\, f^{0 \rho \lambda}{}^{\beta}+f^{\sigma \alpha
  \lambda}{}^{\rho}\, f^{0 \nu \rho}{}^{\beta}
\label{NHfundid}\end{align}
for the structure constants $f^{\sigma\alpha\rho}{}^{\beta}$ of the
3-Lie algebra $\CA_{\rm NH}$, where we introduce the additional constraints
\begin{align}
f^{ijk}{}^l=0 \ , \qquad
f^{ijk}{}^0=\eps^{ijk}
\end{align}
with $i,j,k,l=1,2,3$.
The fundamental identity (\ref{NHfundid}) thus provides the relations
\begin{equation}
f^{0ij}{}^{0}=0 \ , \qquad f^{012}{}^{1}=f^{023}{}^{3} \ , \qquad
f^{012}{}^{2}=-f^{013}{}^{3} \ , \qquad f^{013}{}^{1}=-f^{023}{}^{2} \ ,
\end{equation}
while the other $f^{0ij}{}^k$ remain unconstrained. The structure
constants $f^{0ijk}$ are then proportional to those of a 2-Lie algebra
generated by $\hat x^i$, $i=1,2,3$, according to (\ref{2Liehatx}).

The most general form of the associated Lie algebra $\frg_{\CA_{\rm
    NH}}$ for a Nambu-Heisenberg 3-Lie algebra is thus described by
the commutation relations
\begin{align}
\begin{tabular}{ll}
$[D_{20},D_{30}]=-f^{023}{}^{1}\,D_{10}-f^{023}{}^{2}\,D_{20}-f^{012}{}^{1}\,D_{30}$ \ , &$[D_{12},D_{23}]=D_{20}$ \ , \\[4pt]
$[D_{10},D_{20}]=-f^{012}{}^{1}\,D_{10}+f^{013}{}^{3}\,D_{20}-f^{012}{}^{3}\,D_{30}$ \ , &$[D_{23},D_{13}]=-D_{30}$ \ ,  \\[4pt]
$[D_{10},D_{30}]=f^{023}{}^{2}\,D_{10}-f^{013}{}^{2}\,D_{20}-f^{013}{}^{3}\,D_{30}$ \ ,  &$ [D_{12},D_{13}]=D_{10}$ \ ,  \\[4pt]
$[D_{10},D_{12}]=f^{013}{}^{3}\,D_{12}-f^{012}{}^{3}\,D_{13}$ \ , &$[D_{10},D_{13}]=-f^{013}{}^{2}\,D_{12}-f^{013}{}^{3}\,D_{13}$  \ , \\[4pt]
$[D_{10},D_{23}]=-f^{012}{}^{1}\,D_{13}-f^{023}{}^{2}\,D_{12}$ \ , &$[D_{20},D_{12}]=f^{012}{}^{1}\,D_{12}-f^{012}{}^{3}\,D_{23}$ \ ,  \\[4pt]
$[D_{20},D_{13}]=-f^{013}{}^{3}\,D_{23}-f^{023}{}^{2}\,D_{12} $ \ , &$[D{}^{20},D_{23}]=f^{023}{}^{1}\,D_{12}-f^{012}{}^{1}\,D_{23} $ \ , \\[4pt]
$[D_{30},D_{12}]=-f^{013}{}^{3}\,D_{23}+f^{012}{}^{1}\,D_{13}$ \ , &$[D_{30},D_{13}]=-f^{023}{}^{2}\,D_{13}+f^{013}{}^{2}\,D_{23} $ \ , \\[4pt]
$[D_{30},D_{23}]=f^{023}{}^{1}\,D_{13}+f^{023}{}^{2}\,D_{23}$ \ . & 
\end{tabular}
\end{align}
This is a semisimple Lie algebra with trivial center.
For example, if all structure constants except
$f^{0ij}{}^k=\eps^{ijk}$ are set to zero, which is consistent with the
fundamental identity, then the 3-Lie algebra $\CA_{\rm NH}$ is
isomorphic to ${A}_4$, whose associated Lie algebra is $\mathfrak{so}(4)$.

\bibliographystyle{latexeu}

\bibliography{bigone}

\end{document}